\documentclass{article}
\usepackage{spconf,amsmath,graphicx,hyperref}

\usepackage{color}
\usepackage{multirow}
\usepackage{comment}
\usepackage{array}
\usepackage{enumerate}
\usepackage{url}
\usepackage{tabularx}
\usepackage{setspace}
\usepackage{float}
\usepackage{rotating}

\newlength\savedwidth
\newcommand{\wcline}[1]{\noalign{\global\savedwidth\arrayrulewidth\global\arrayrulewidth 1.0pt} \cline{#1}
\noalign{\global\arrayrulewidth\savedwidth}}
\renewcommand\footnotesize{\scriptsize}

\title{Joint Analysis of Acoustic Scenes and Sound Events Based on Semi-Supervised Training of Sound Events With Partial Labels}

\name{Keisuke Imoto}

\address{
Graduate School of Informatics, Kyoto University, Japan \\
\texttt{keisuke.imoto@ieee.org}
}

\begin{document}
\ninept
\maketitle
%
%---------------------------------------------------
\begin{abstract}
%---------------------------------------------------
Annotating time boundaries of sound events is labor-intensive, limiting the scalability of strongly supervised learning in audio detection. To reduce annotation costs, weakly-supervised learning with only clip-level labels has been widely adopted. As an alternative, partial label learning offers a cost-effective approach, where a set of possible labels is provided instead of exact weak annotations. However, partial label learning for audio analysis remains largely unexplored. Motivated by the observation that acoustic scenes provide contextual information for constructing a set of possible sound events, we utilize acoustic scene information to construct partial labels of sound events. On the basis of this idea, in this paper, we propose a multitask learning framework that jointly performs acoustic scene classification and sound event detection with partial labels of sound events. While reducing annotation costs, weakly-supervised and partial label learning often suffer from decreased detection performance due to lacking the precise event set and their temporal annotations. To better balance between annotation cost and detection performance, we also explore a semi-supervised framework that leverages both strong and partial labels. Moreover, to refine partial labels and achieve better model training, we propose a label refinement method based on self-distillation for the proposed approach with partial labels.%---------------------------------------------------
\end{abstract}
%---------------------------------------------------
%

%---------------------------------------------------
%\vspace{0pt}
\section{Introduction}
\label{sec:intro}
%\vspace{0pt}
%---------------------------------------------------
%
Computational analysis of environmental sounds has recently attracted much attention in the field of acoustic signal and speech processing.
Environmental sound analysis, which is not limited to speech or musical sound analysis, greatly expands the range of sound-based applications, such as media retrieval, hearing aids, machine condition monitoring, self-driving cars, robot auditions, and biomonitoring systems (\cite{Fonseca_DCASE2018_01,Nishida_EUSIPCO2022_01,Morfi_JASA2021_01}).

In environmental sound analysis, acoustic scene classification (ASC) and sound event detection (SED) are fundamental tasks.
Of these tasks, ASC estimates an acoustic scene label most related to an input sound.
SED predicts all sound event labels and their corresponding start and end times in an input sound.
Recently, various ASC and SED methods based on neural networks have been adopted, and they effected a remarkable improvement in performance.
For example, Valenti et al. (\cite{Valenti_IJCNN2017_01}) and Ford et al. (\cite{Ford_INTERSPEECH2019_01}) have proposed ASC systems using the convolutional neural network (CNN) and ResNet, respectively.
Kong et al. proposed an ASC method using a pre-trained model with a large-scale audio dataset (\cite{Kong_TASLP2020_01}).
\c{C}ak\i r et al. introduced a SED technique incorporating a convolutional recurrent neural network (CRNN) (\cite{Cakir_TASLP2017_01}). 
More recently, Kong et al. (\cite{Kong_TASLP2020_02}) and Miyazaki et al. (\cite{Miyazaki_DCASE2020_01}) proposed Transformer- and Conformer-based SED methods, respectively, which have been widely employed in many studies.

Acoustic scenes and sound events are mutually related and they are effectively estimated by utilizing mutual information.
For instance, in the acoustic scene \textit{home}, the sound events \textit{cutlery} and \textit{door opening/closing} tend to occur, whereas the sound events \textit{car} and \textit{bird singing} are not likely to occur.
Taking into account the relationship between acoustic scenes and sound events, Mesaros et al. (\cite{Mesaros_EUSIPCO2011_01}) proposed a SED method leveraging knowledge on acoustic scenes.
Similarly, Imoto and colleagues (\cite{Imoto_TASLP2019_01}) and Hou et al. (\cite{Hou_SLP2023_01}) proposed ASC methods that take into account the association between acoustic scenes and sound events, through the use of Bayesian generative models and graph neural network, respectively.
In more recent works, Bear et al. (\cite{Bear_INTERSPEECH2019_01}), Tonami et al. (\cite{Tonami_WASPAA2019_01}), and Jung et al. (\cite{Jung_ICASSP2021_01}) proposed the joint analysis of acoustic scenes and sound events using the multitask learning (MTL) framework of ASC and SED, which learns both tasks simultaneously.

\begin{figure}[t!]
\centering
%\vspace{0pt}
\includegraphics[width=0.9\columnwidth]{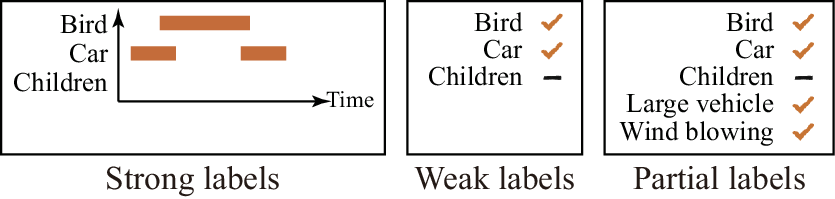}
%\vspace{0pt}
\caption{Illustration comparing strong, weak, and partial labels in sound event. Strong labels provide sound event classes and their time stamps, weak labels indicate which event classes occur within an audio clip, and partial labels provide a candidate set of event labels.}
\label{fig:label_01}
%\vspace{0pt}
\end{figure}
%

%
%\vspace{0pt}
\begin{figure*}[t!]
\centering
%\vspace{0pt}
\begin{tabular}{c}
\begin{minipage}[t!]{0.40\linewidth}
\centering
\includegraphics[width=0.944\columnwidth]{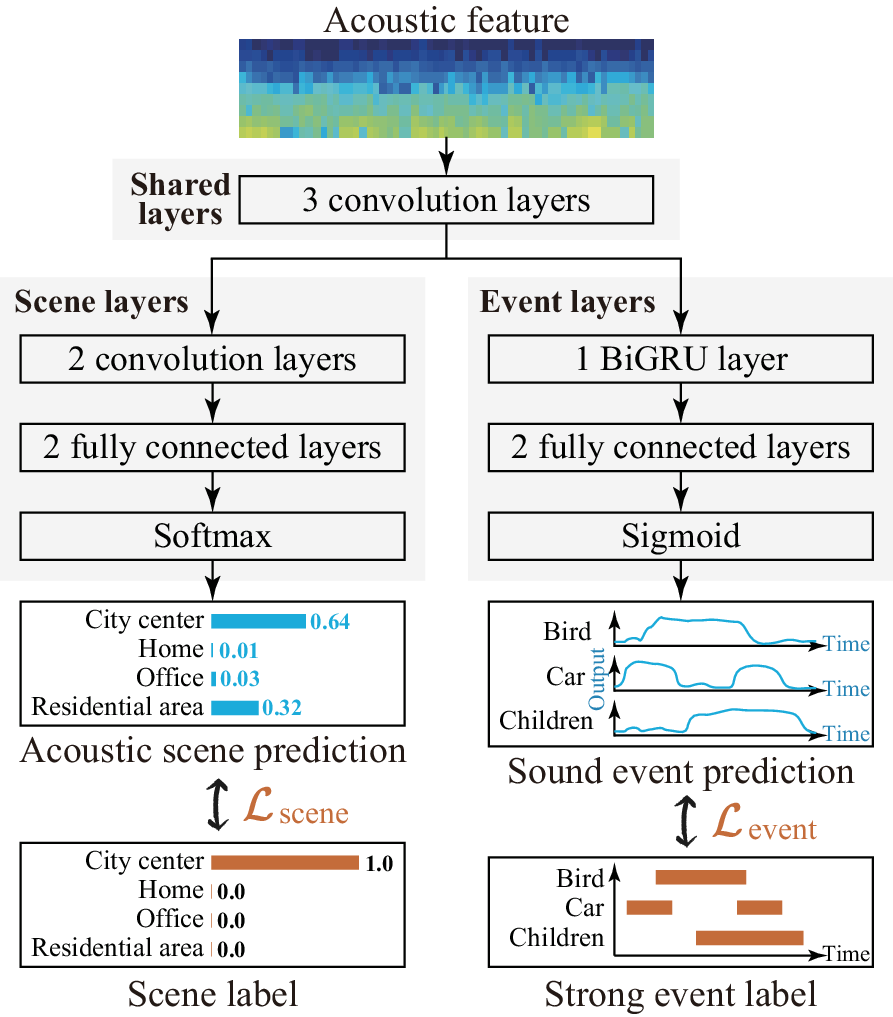}
%\vspace{0pt}
\caption{Network structure of conventional MTL-based method (\cite{Tonami_WASPAA2019_01})}
\label{fig:conventionalMTL}
\end{minipage}
\hspace{8pt}
\begin{minipage}[t!]{0.57\linewidth}
%\vspace{0pt}
\centering
\includegraphics[width=0.96\columnwidth]{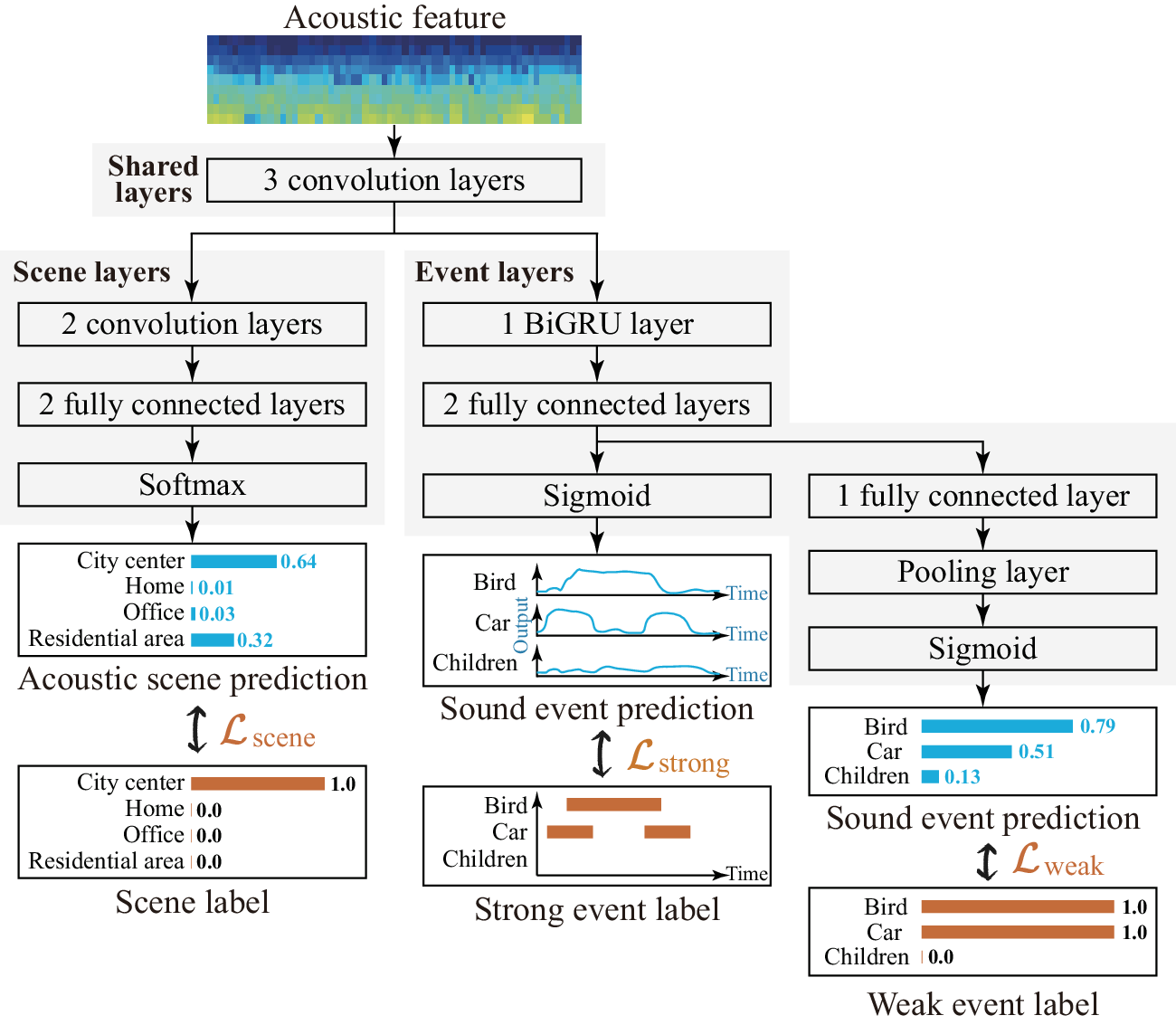}
%\vspace{0pt}
\caption{Network structure of MTL-based method using weak labels of sound events}
\label{fig:weakMTL}
\end{minipage}
\end{tabular}
%\vspace{0pt}
\end{figure*}

Many methods for environmental sound analysis are based on the supervised learning scheme, which train model parameters using large-scale strongly annotated data.
However, annotating labels for environmental sounds, especially annotating time boundaries of sound events, is very laborious.
Moreover, there are potential applications where collecting large-scale annotated data itself is difficult. For example, in-home monitoring systems must address privacy concerns, making it difficult to share audio data with unspecified annotators. Similarly, in ecological monitoring, expert knowledge is required to annotate species-specific sounds such as bird calls or amphibian vocalizations, limiting the scalability of manual annotation.
To mitigate this challenge, in the context of single-task SED, many methods using weakly-supervised learning have been proposed (\cite{Kumar_ACMMM2016_01, Turpault_DCASE2019_01}).
In the paradigm of weakly-supervised learning for SED, only information on clip-level activations of sound events is provided in the training stage, whereas sound event labels and their time boundaries are estimated in the inference stage.
For the joint analysis of acoustic scenes and sound events, Tsubaki et al. (\cite{Tsubaki_IWAENC2022_01}) and Igarashi et al. (\cite{Igarashi_APSIPA2023_01}) proposed a method that applies the weakly-supervised SED approach to the MTL framework.

To further reduce the cost of annotation, partial label learning, in which a detection or classification model is trained using a set of possible labels, has also been proposed in image analysis (\cite{Cour_JMLR2011_01}).
However, partial label learning for SED has been largely unexplored.
To clarify the differences among strong, weak, and partial labels in SED, Fig.~\ref{fig:label_01} illustrates each labeling scheme.
The partial labels actually are a form of weak labels; however, unlike conventional weak labels that only include true sound event classes, partial labels represent a set of possible sound event classes, which may contain additional labels beyond the ground truth.
Although annotating precise weak labels still requires substantial effort, annotating only a set of possible labels can significantly reduce annotation costs and facilitate the creation of large-scale training data.
Note that the conventional method of partial label learning (\cite{Cour_JMLR2011_01}) has addressed the image classification task, where exactly one true label is assumed to exist in each possible label set. In contrast, our work focuses on SED, where multiple true sound event labels may be present within the partial labels.

In environmental sound analysis, the strong correlation between acoustic scenes and sound events enables the generation of effective partial labels of sound events, as scene information can be used to constrain the candidate set of sound event classes.
Since annotating acoustic scene labels requires much less effort than annotating precise weak labels of sound events, leveraging scene information offers a cost-effective way to guide SED model training.
Thus, in this work, we propose the MTL framework of SED and ASC using partial labels of sound events, which offers a suitable and practical setting for exploring the use of partial label learning in environmental sound analysis.
On the other hand, introducing partial label learning may result in performance degradation compared with methods using strong labels.
Thus, we further explore an MTL-based joint analysis of acoustic scenes and sound events using both strong and partial labels of sound events simultaneously, that is, a semi-supervised approach.
We then evaluate the performances of ASC and SED in detail and characterize the behavior of the proposed method with partial labels.

The subsequent sections of this paper are as follows.
In Section 2, we discuss conventional methodologies for ASC, SED, and the joint analysis of acoustic scenes and sound events by leveraging multitask learning.
Section 3 is dedicated to introducing our methods of joint analysis of acoustic scenes and sound events utilizing semi-supervised approaches with partial labels of sound events.
The evaluation experiments to validate the detailed performance of scene classification and event detection are presented in Section 4.
Finally, in Section 5, we conclude this paper and discuss potential directions for a future work.
%
%
%---------------------------------------------------
%\vspace{0pt}
\section{Conventional Methods}
\label{sec:conventional}
%\vspace{0pt}
%---------------------------------------------------
%- - - - - - - - - - - - - - - - - - - - - - - - - - -
%\vspace{0pt}
\subsection{Acoustic Scene Classification and Event Detection}
\label{ssec:ConvASCandSED}
%\vspace{0pt}
%- - - - - - - - - - - - - - - - - - - - - - - - - - -
In this section, we overview basic implementations for ASC and SED using neural networks.
Many systems for ASC and SED first extract a time--frequency representation of the acoustic signal $\textbf{X} \in \mathcal{R}^{D \times T}$ from an audio input.
Here, $D$ and $T$ represent the numbers of frequency bins and time frames, respectively.
The log mel-band spectrogram or a time series of mel frequency cepstrum coefficients (MFCCs) is typically used for the acoustic feature.
The extracted acoustic feature is subsequently fed to the ASC or SED networks, which calculate logits $\textbf{y}$ for classifying acoustic scenes or detecting sound events, respectively.

\begin{table*}[t!]
\small
\caption{Partial labels generated by ChatGPT o3-mini-high for TUT Acoustic Scenes 2016 and TUT Sound Events 2016/2017}
%\vspace{-5pt}
\label{tbl:event_occurrence_partial}
\centering
\begin{tabular}{c}
\includegraphics[width=2.0\columnwidth]{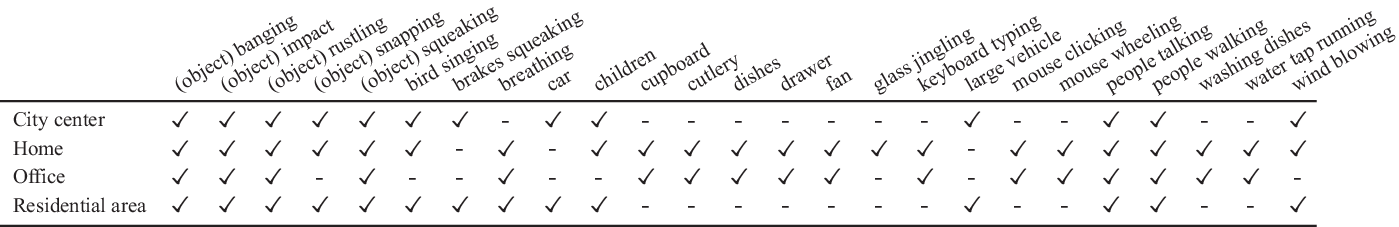}
\end{tabular}
%\vspace{0pt}
\end{table*}
\begin{table*}[t!]
\small
\caption{Sound event list of training dataset in TUT Acoustic Scenes 2016 and TUT Sound Events 2016/2017}
\label{tbl:event_occurrence}
%\vspace{-5pt}
\centering
\begin{tabular}{c}
\includegraphics[width=2.0\columnwidth]{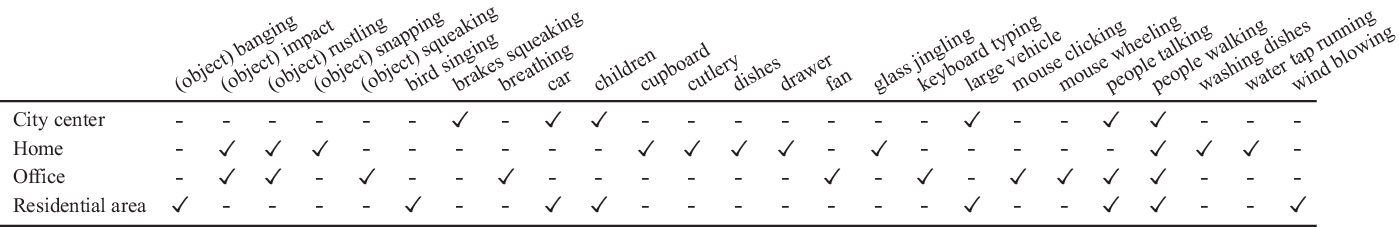}
\end{tabular}
%\vspace{0pt}
\end{table*}

As for the ASC network, the model parameters are trained using the logits and the cross-entropy (CE) loss function ${\mathcal L}_{{\mathsf{scene}}}$ as

%
%\vspace{0pt}
\begin{align}
{\mathcal L}_{{\mathsf{scene}}} &= - \sum^{N}_{n=1} {\Big \{} z_{n} \log \bigl( \sigma \left( y_{n} \right) \bigr) {\Big \}},
\label{eq:scene_loss}
%\vspace{0pt}
\end{align}
%\vspace{0pt}
%

\noindent where $N$, $z_{n} \in \{0,1\}$, and $\sigma$ are the number of acoustic scene classes, the acoustic scene label, and softmax function, respectively.

On the other hand, the parameters of the SED network are tuned using the following binary cross-entropy (BCE) loss function as follows:

%
%\vspace{0pt}
\begin{align}
\hspace{-3pt} {\mathcal L}_{{\mathsf{event}}} &= - \hspace{-1.5pt} \sum^{T}_{t=1} \hspace{-1pt} {\Big \{} {\bf z}_{t} \log ( {\bf y}_{t} ) + (1 \! - \! {\bf z}_{t}) \log \bigl(1 \! - \! s ( {\bf y}_{t} ) \bigr) \hspace{-1pt} {\Big \}} \nonumber\\[1pt]
&\hspace{0pt} = - \!\! \hspace{-1.8pt} \sum^{T\!, \hspace{1pt} M}_{t,m=1} \hspace{-4.5pt} {\Big \{} \hspace{-0.5pt} z_{t,m} \log s ( y_{t,m} ) \hspace{-1pt} \! + \! (1 \! - \! z_{t,m}) \hspace{-0.8pt} \log \bigl( 1 \! - \! s ( y_{t,m}) \hspace{-1pt} \bigr) \hspace{-2pt} {\Big \}} \hspace{-0.3pt} ,\hspace{-4pt}
%}
\label{eq:event_loss}
%\vspace{0pt}
\end{align}
%\vspace{0pt}
%

\noindent where $T$, $M$, $z_{t,m}$, and $s$ indicate the number of time frames in a sound clip, the number of sound event classes, the target event label in the time frame $t$ for the sound event $m$, and the sigmoid function, respectively.
%
%
%- - - - - - - - - - - - - - - - - - - - - - - - - - -
%\vspace{0pt}
\subsection{Joint Analysis of Acoustic Scenes and Sound Events Based on Multitask Learning}
\label{ssec:ConvMTL}
%\vspace{0pt}
%- - - - - - - - - - - - - - - - - - - - - - - - - - -
In the realm of environmental sound analysis, numerous methods address scene classification and event detection as individual tasks.
Only a few works focus on the idea that information on acoustic scenes and sound events mutually enhances the performance in ASC and SED, and methods that jointly analyze acoustic scenes and sound events, has been proposed (\cite{Bear_INTERSPEECH2019_01,Tonami_WASPAA2019_01,Jung_ICASSP2021_01}).

A typical implementation of the joint analysis of acoustic scenes and sound events utilizes an MTL-based neural network, which shares part of the network and information on acoustic scenes and sound events, as shown in Fig.~\ref{fig:conventionalMTL}.
The conventional method first extracts a feature embedding common to acoustic scenes and sound events in the shared layers.
The resultant feature embedding is subsequently fed to the dedicated layers tailored for ASC and SED.
For the dedicated layers for acoustic scenes and sound events, CNN, the recurrent neural network (RNN), and the Transformer encoder are often employed.

To train the model parameters, the conventional methods (\cite{Tonami_WASPAA2019_01}) adopt a loss function represented by the linear combination of Eqs. (1) and (2) with acoustic scene labels and strong sound event labels.
%
%\vspace{0pt}
\begin{align}
{\mathcal L} &= \alpha {\mathcal L}_{{\mathsf{scene}}} + \beta {\mathcal L}_{{\mathsf{event}}}
\label{eq:mtl_loss}
%\vspace{0pt}
\end{align}
%\vspace{0pt}
%

\noindent Here, $\alpha$ and $\beta$ are the constant weights for ASC and SED losses, respectively.
In this paper, we set $\beta = 1.0$ without loss of generality.
\begin{figure*}[t!]
\centering
%\hspace*{-3pt}
%\vspace{0pt}
\includegraphics[width=1.88\columnwidth]{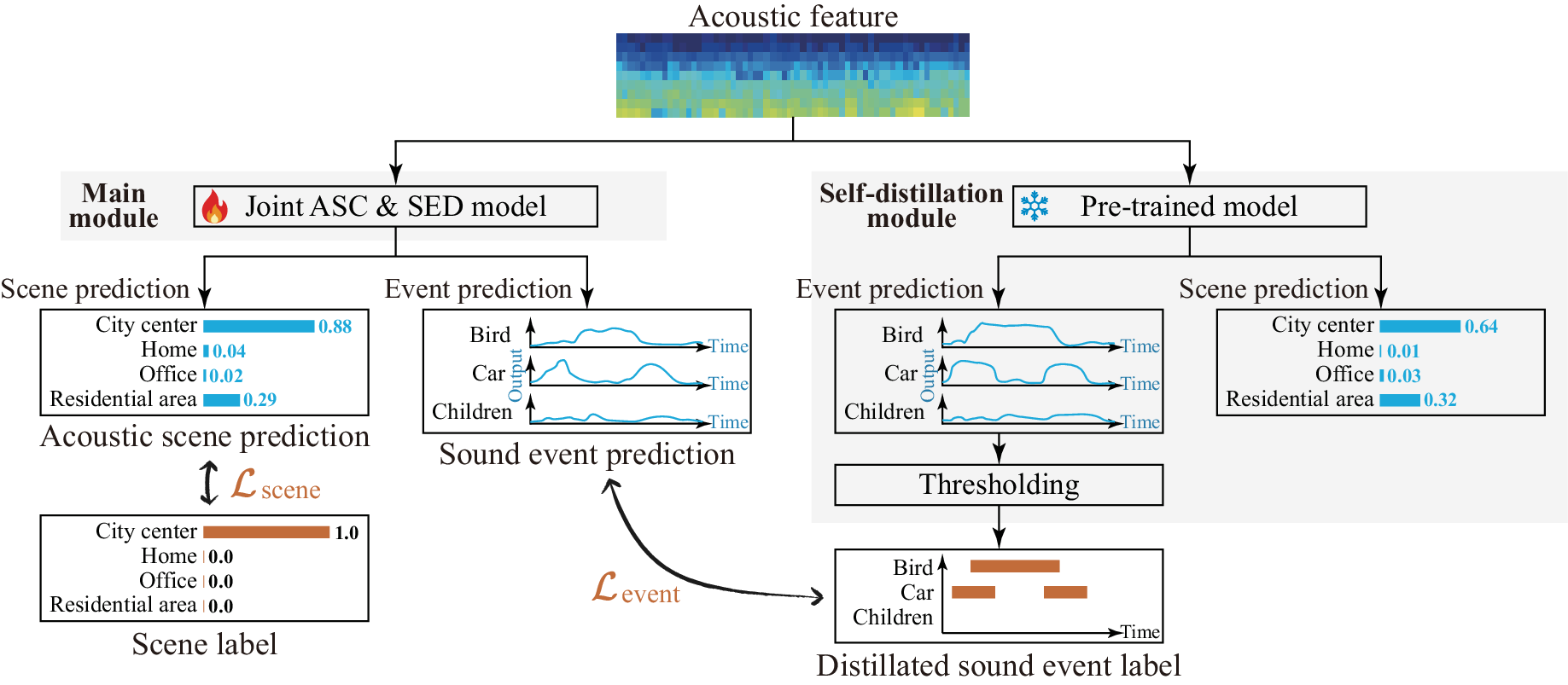}
\vspace{-3pt}
\caption{Self-distillation-based model training for semi-supervised method using partial labels of sound events}
\label{fig:self_distill_01}
\vspace{5pt}
\end{figure*}
%
%
%---------------------------------------------------
%\vspace{-0pt}
\subsection{Weakly-Supervised Method for Joint Analysis of Acoustic Scenes and Sound Events}
\label{ssec:WeakMTL}
%\vspace{-0pt}
%---------------------------------------------------
Annotating time boundaries of sound events is labor-intensive and time-consuming.
To overcome the challenge of annotating strong labels of sound events, in the single-task SED scenarios, many methods apply a weakly-supervised scheme using weak labels of sound events.
Here, weak labels of sound events only have information on the presence or absence of sound events in a sound clip.
Many weakly-supervised methods for single-task SED employ the multiple-instance learning (MIL) framework (\cite{Dietterich_AI1997_01,Wang_ICASSP2019_01}), which makes a final decision by aggregating small bag-level decisions.
In the case of SED, the system outputs a clip-level detection result of a sound event by aggregating frame-level decisions.

Tsubaki et al. (\cite{Tsubaki_IWAENC2022_01}) proposed a framework for the joint analysis of ASC and SED, in which weakly-supervised learning is integrated into the SED task.
Figure~\ref{fig:weakMTL} shows the network structure of the conventional method (\cite{Tsubaki_IWAENC2022_01}), which has two branches in the event layers.
One branch has the pooling layer corresponding to the MIL framework and enables the weakly-supervised training in SED.
The other branch only has the sigmoid function to hold temporal information and it enables us to estimate time stamps of sound events in the inference stage.

To train the model parameters of the weakly-supervised method, the linear combination of the ASC and SED losses represented by Eq.~(\ref{eq:mtl_loss}) are also used, whereas we modify the SED loss $\mathcal{L}_{\mathsf{event}}$ as follows:

%
%\vspace{0pt}
\begin{align}
&\hspace{-2pt} \mathcal{L}_{\mathsf{event}} = \gamma \mathcal{L}_{\mathsf{strong}} + \zeta \mathcal{L}_{\mathsf{weak}}\nonumber\\[2pt]
&\hspace{2pt} = - \gamma \hspace{-1pt} \sum_{m,t=1}^{M\hspace{-0.8pt}, \hspace{0.8pt}T} \hspace{-2pt} {\Big \{} z_{m,t} \log s(y_{m,t}) \hspace{-1pt} + \hspace{-1pt} (1 \hspace{-1pt} - \hspace{-1pt} z_{m,t}) \log {\big (} 1 \hspace{-1pt} - \hspace{-1pt} s(y_{m,t}) {\big )} \hspace{-1pt} {\Big \}}\nonumber\\[0pt]
&\hspace{12pt} - \zeta \hspace{-1pt} \sum_{m=1}^{M} \hspace{-2pt} {\Big \{} z_{m} \log s(y_{m}) \hspace{-1pt} + \hspace{-1pt} (1-z_{m}) \log {\big (} 1 \hspace{-1pt} - \hspace{-1pt} s(y_{m}) {\big )} \hspace{-1pt} {\Big \}},\nonumber\\[-6pt]
\label{eq:weak_loss}
\end{align}
%\vspace{0pt}
%

\noindent where $z_{m}$ and $y_{m}$ are the weak label for the sound event $m$ and the clip-level prediction of the sound event $m$, respectively.
$\gamma$ and $\zeta$ are the constant weights for the losses for the frame- and clip-level predictions, respectively.
Here, in the training stage, the strong event label $z_{m,t}$ is prepared from the weak label $z_{m}$ as a pseudo-sound event label as

%\vspace{0pt}
\begin{align}
\overbrace{\hspace{10.5em}}^{T} \hspace*{17pt} \nonumber\\[-5pt]
\mathbf{Z}_{\mathsf{pseudo\_strong}} =
\left(\begin{array}{ccccc}
z_{1} & \cdots & z_{1} & \cdots & z_{1} \\
\vdots & \ddots & \vdots && \vdots \\
z_{m} & \cdots & z_{m} & \cdots & z_{m} \\
\vdots && \vdots & \ddots & \vdots \\
z_{M} & \cdots & z_{M} & \cdots & z_{M}
\end{array}\right).
\label{pseudo_strong}
\end{align}
%\vspace{0pt}
%
%
%---------------------------------------------------
%\vspace{-0pt}
\subsection{Semi-Supervised Method for Joint Analysis of Acoustic Scenes and Sound Events With Weak Labels of Sound Events}
\label{ssec:SemiMTL_Weak}
%\vspace{-0pt}
%---------------------------------------------------
The MTL-based method using weak labels of sound events mitigates the challenge inherent in collecting strong labels of sound events to some extent.
Nonetheless, joint analysis of ASC and SED using weak labels results in a lower SED performance as compared with that using strongly labeled data.
To address this problem, a semi-supervised learning scheme has been proposed for SED, in the context of joint analysis of ASC and SED (\cite{Igarashi_APSIPA2023_01}).
In general, methods that amalgamate both supervised and unsupervised training modalities are termed as semi-supervised learning.
In this paper, however, we specifically refer to the method leveraging both strong and weak/partial labels for the SED model training as semi-supervised learning.

Let the acoustic feature sets with strong and weak labels be $\mathcal{X}_{\mathsf{strong}}$ and $\mathcal{X}_{\mathsf{weak}}$, respectively.
Similarly, we consider that the strong and weak label sets as $\mathcal{Z}_{\mathsf{strong}}$ and $\mathcal{Z}_{\mathsf{weak}}$, respectively.
For the semi-supervised approach, we construct the acoustic feature and label sets as

%\vspace{0pt}
\begin{align}
\mathcal{X}_{\mathsf{semi}} = \{\mathcal{X}_{\mathsf{strong}},\mathcal{X}_{\mathsf{weak}}\},\\
\mathcal{Z}_{\mathsf{semi}} = \{\mathcal{Z}_{\mathsf{strong}},\mathcal{Z}_{\mathsf{weak}}\}.
\end{align}
\vspace{0pt}

For the semi-supervised approach, we can employ various network architectures once it is designed with four key modules: (i) an acoustic embedding extractor, (ii) acoustic scene classifier, and (iii)(iv) sound event detectors with weak labels and strong labels.
In this paper, we illustrate the conventional semi-supervised method with the same network structure as that of the weakly-supervised method as shown in Fig.~\ref{fig:weakMTL}.

To train the model parameters, Eqs. (\ref{eq:mtl_loss}) and (\ref{eq:weak_loss}) are also used as the loss function, while $\gamma$ and $\zeta$ are replaced with the following Kronecker delta functions:

\begin{align}
\delta_{\gamma} \ =\ 
\begin{cases}
1 & \textrm{if} \ {\bf z} \ \textrm{is strong label}\\
0 & \textrm{otherwise},
\end{cases}
\end{align}

\begin{align}
\delta_{\zeta} \ =\ 
\begin{cases}
1 & \textrm{if} \ {\bf z} \ \textrm{is weak label}\\
0 & \textrm{otherwise}.
\end{cases}
\end{align}
\vspace{0pt}

\noindent This strategy enables switching between SED networks depending on whether strong labels are available or only weak labels can be used.
The semi-supervised method using strong and weak labels is expected to achieve more reliable model training than the weakly-supervised method that relies on pseudo labels of sound events.
%
%---------------------------------------------------
%\vspace{-0pt}
\section{Joint Analysis of Acoustic Scenes and Sound Events Based on Semi-Supervised Approach With Partial Labels of Sound Events}
\label{sec:proposed}
%\vspace{-0pt}
%---------------------------------------------------
Annotating weak labels for sound events indeed alleviates the cost of labor involved in annotating strong labels for sound events.
However, compared with annotating acoustic scene labels, annotating weak labels for sound events is still labor-intensive.
Therefore, we propose a method that utilizes acoustic scene labels to generate candidate weak labels for sound events, which can then be employed as partial labels in model training.
In particular, this paper explores the use of partial labels in semi-supervised learning for joint analysis of acoustic scenes and sound events.

To generate partial labels of sound events, we can utilize acoustic scene labels in several ways: one approach is to pre-construct candidate label lists for each acoustic scene, while an alternative is to generate these candidate lists using a pre-trained model, such as a large language model (LLM).
For instance, in our experiments in this study, we created partial weak labels by inputting acoustic scene labels into ChatGPT o3-mini-high\footnote{The label list was generated using ChatGPT o3-mini-high on February 02, 2025}\footnote{We have also generated partial labels using the same prompts with several LLMs, including ChatGPT 5 Thinking and Gemini 2.5 Pro. These models produced sound event label sets largely similar to those obtained with ChatGPT o3-mini-high.}.
The prompt used for generating the partial labels is provided in Appendix, and the resulting constructed partial labels are listed in Table~\ref{tbl:event_occurrence_partial}.
Compared with the actual sound event label list shown in Table~\ref{tbl:event_occurrence}, the generated partial label set includes a significantly larger number of candidate sound events, such as generic events like \textit{(object) impact}, which commonly appear in various acoustic scenes. On the other hand, we observed no case where actually occurring events were omitted from the generated labels. Given that the partial labels were created using the publicly available LLM, we believe that the quality of partial labels reflects a realistic application scenario, and their reliability is sufficient for practical use.

In this work, we further apply a method that refines sound event labels and generates pseudo strong labels using self-distillation, to mitigate the noise in the partial label set generated using an LLM.
This self-distillation-based approach represents one of the simplest methods for label refinement in semi-supervised learning. To verify the feasibility of model training from partial labels in the multitask learning of sound events and acoustic scenes, we employ this simple label refinement method in this study.
The procedure for this partial label learning is shown in Fig.~\ref{fig:self_distill_01}.
First, partial labels are treated as weak ground truth labels, and the joint ASC and SED model is trained using both strong and partial labels according to the method described in Section II-D.
Once the model parameters have been trained, the pre-trained model is frozen and the training data with partial labels is fed into the self-distillation module to obtain logits.
The posterior probabilities of the sound events are then calculated using a sigmoid function, and the distillated strong event labels are obtained by thresholding them with $\phi$.
After that, the main module is re-trained using the strong and distillated strong event labels with the conventional MTL-based method described in Section II-B.

\begin{table}[t]
%\vspace{0pt}
\small
\caption{Detailed structure of MTL network of ASC and SED using weak/partial labels}
\vspace{5pt}
\label{tbl:networks}
\centering
\begin{tabular}{ccccc}
\wcline{1-5}
&\\[-9pt]
\multicolumn{5}{c}{\textbf{Shared layers}}\\
\wcline{1-5}
&\\[-9pt]
\multicolumn{5}{c}{Log-mel energy (500 frames $\times$ 64 mel bin)}\\[0pt]
\cline{1-5}
\!&\\[-9pt]
\multicolumn{5}{c}{3$\times$3 kernel size/128 ch.}\\[-1pt]
\multicolumn{5}{c}{Batch norm., Leaky ReLU}\\[-1pt]
\multicolumn{5}{c}{1$\times$8 Max pooling}\\[0pt]
\cline{1-5}
\!&\\[-9pt]
\multicolumn{5}{c}{$\begin{pmatrix} \textrm{3$\times$3 kernel size/128 ch.}\\[-1pt]
\textrm{Batch norm., Leaky ReLU}\\[-1pt]
\textrm{1$\times$2 Max pooling}
\end{pmatrix}$ $\times$ 2
}\\[0pt]
\!&\\[-9pt]
\wcline{1-5}
\multicolumn{1}{c}{}\\[-9pt]
\multicolumn{1}{c}{\!\!\!\textbf{Scene layers}}\!\!\!&\!\!\!\!&\multicolumn{3}{c}{\textbf{Event layers}}\\
\wcline{1-5}
\multicolumn{1}{c}{}\\[-9pt]
\multicolumn{1}{c}{\!\!\!3$\times$3 kernel size/256 ch.}\!\!\!\!&\!\!\!\!&\\[-1pt]
\multicolumn{1}{c}{\!\!\!Batch norm., Leaky ReLU}\!\!\!\!&\!\!\!\!&\multicolumn{3}{c}{\!\!Transformer Enc. w/ 512 units}\\[-1pt]
\multicolumn{1}{c}{\!\!\!25$\times$1 Max pooling}\!\!\!\!&\!\!\!\!&\\
\cline{1-1} \cline{3-5}
\multicolumn{1}{c}{}\\[-9pt]
\multicolumn{1}{c}{\!\!\!3$\times$3 kernel size/256 ch.}\!\!\!\!&\!\!\!\!&\multicolumn{3}{c}{\!\!\multirow{3}{*}{FC w/ 48 units, Leaky ReLU}}\\[-1pt]
\multicolumn{1}{c}{\!\!\!Batch norm., Leaky ReLU}\!\!\!\!&\!\!\!\!&\\[-1pt]
\multicolumn{1}{c}{\!\!\!Global max pooling}\!\!\!\!&\!\!\!\!&\multicolumn{3}{c}{\!\!}\\
\cline{1-1} \cline{3-5}
\multicolumn{1}{c}{}\\[-9pt]
\multicolumn{1}{c}{\!\!\!FC w/ 32 units, Leaky ReLU}\!\!\!\!&\!\!\!\!&\multicolumn{3}{c}{\!\!FC w/ 25 units}\\[0pt]
\cline{1-1}\cline{3-5}
\multicolumn{1}{c}{}\\[-9pt]
\multicolumn{1}{c}{\!\!\!FC w/ 4 units, Softmax}\!\!\!\!&\!\!\!\!&\!\!\!Sigmoid\!\!&\!\!\!\!&\!\!\!FC w/ 16 units\!\!\!\\
\wcline{1-1}\wcline{3-3}
\multicolumn{1}{c}{}\\[-9pt]
\multicolumn{1}{c}{}&\!\!\!\!&&\!\!\!\!&\!\!\!Leaky ReLU\!\!\!\\
\cline{5-5}
\multicolumn{1}{c}{}\\[-9pt]
\multicolumn{1}{c}{}&\!\!\!\!&&\!\!\!\!&\!\!\!Global max pooling\!\!\!\\
\cline{5-5}
\multicolumn{1}{c}{}\\[-9pt]
\multicolumn{1}{c}{}&\!\!\!\!&&\!\!\!\!&\!\!\!Sigmoid\!\!\!\\
\wcline{5-5}
\end{tabular}
%\vspace{0pt}
\end{table}
\begin{table}[t]
%\vspace{0pt}
\footnotesize
\centering
\caption{Experimental conditions}
\vspace{3pt}
\label{tbl:parameter}
\begin{tabular}{ll}
\wcline{1-2}
&\\[-7pt]
Acoustic feature & Log-mel energy (64 dim.)\\
Frame length/shift & 40 ms/20 ms\\
Length of sound clip & 10 s\\
Optimizer & RAdam (\cite{Liu_ICLR2020_01})\\[0pt]
SED detection threshold & 0.5\\[0pt]
$\alpha$, $\beta$, $\gamma$, $\zeta$ & 0.001, 1.0, 1.0, 0.01\\[0pt]
$\rho_{_{GTC}}$, \ $\rho_{_{DTC}}$ & 0.1, 0.1\\[0pt]
Threshold $\phi$ for self-distillation & 0.2\\[1pt]
\wcline{1-2}
\end{tabular}
\vspace{10pt}
\end{table}
%
%
%-----------------------------------------------------
%\vspace{0pt}
\section{Evaluation Experiments}
\label{sec:experiment}
%\vspace{0pt}
%-----------------------------------------------------
%- - - - - - - - - - - - - - - - - - - - - - - - - - - -
%\vspace{-0pt}
\subsection{Experimental Conditions}
\label{ssec:condition}
%\vspace{0pt}
%- - - - - - - - - - - - - - - - - - - - - - - - - - - -
We carried out experiments to evaluate the conventional and proposed MTL-based joint analyses of acoustic scenes and sound events.
For the evaluation experiments, we constructed a dataset composed of the TUT Acoustic Scene 2016/2017 and TUT Sound Events 2016/2017 (\cite{Mesaros_EUSIPCO2016_01,Mesaros_DCASE2017_01}), which includes four acoustic scenes (\textit{city center}, \textit{home}, \textit{office}, and \textit{residential area}) and 25 sound events (e.g., \textit{bird singing}, \textit{car}, \textit{dishes}, and \textit{keyboard typing}).
The dataset contains a total of 266 min of sounds, which includes 192 min of sounds for model training and 74 min of sounds for evaluation.
The partial labels were created using ChatGPT o3-mini-high, which was one of the most capable and generally applicable models available at the time of our experiments.
All experiments were conducted on a single Intel Xeon Gold 6128 Processor and an NVIDIA RTX 6000 Ada Generation GPU.
The details of the dataset and baseline code are available\footnote{\url{https://www.ksuke.net/dataset/}}\footnote{\url{https://github.com/KeisukeImoto/mtl_sed_asc}}.

We calculated the 64-dimensional log mel-band spectrogram with a frame length of 40 ms and a hop size of 20 ms.
The model structure used for our experiment is shown in Figs.~\ref{fig:conventionalMTL}, \ref{fig:weakMTL}, and \ref{fig:self_distill_01}, and Table~\ref{tbl:networks}, which are based on conventional works (\cite{Tonami_WASPAA2019_01}).
In our preliminary experiments, we also evaluated other sophisticated model architectures for the SED-specific layers such as the Transformer and Conformer.
However, these model architectures showed performance nearly equivalent to that of the CRNN-based method.
This may be because we used the dataset with limited size.
In this study, we thus adopt the same model architecture as in previous research to enable direct comparisons.
The threshold $\phi$ for self-distillation was determined through the preliminary experiment using cross-validation setup on the training data as shown in Table~\ref{tbl:parameter}.
The other experimental conditions are also found in Table~\ref{tbl:parameter}.
These settings and hyperparameters were determined by referring to (\cite{Tonami_WASPAA2019_01}).
Since the original dataset has strong labels of sound events, we randomly selected samples from the training set and discarded time stamps to create weak labels.
We conducted the evaluation experiments 10 times for each experimental condition with random initial values of model parameters.
\begin{table}[t!]
%\vspace{0pt}
\scriptsize
\caption{Overall performance characteristics of ASC and SED. We conducted the experiments with 30\% of the strongly labeled data and 70\% of the weakl/partial labeled data under the semi-MTL condition.}
\vspace{5pt}
\hspace*{-8pt}
\label{tbl:performance01}
\centering
\begin{tabular}{ccccccccccc}
\wcline{1-9}
&\\[-6pt]
\multicolumn{1}{c}{\!\!\!\multirow{4}{*}{\textbf{Method}}}&\multicolumn{2}{c}{\!\!\!\multirow{2}{*}{\textbf{Scene}}}&\!\!\!\!\!\!&\multicolumn{2}{c}{\textbf{Event}}&\!\!\!\!\!\!&\multicolumn{2}{c}{\!\!\!\textbf{Event}}\\[-1pt]
&&&\!\!\!\!\!\!&\multicolumn{2}{c}{\!\!\!\textbf{(Segment-based)}}&\!\!\!\!\!\!&\multicolumn{2}{c}{\!\!\!\textbf{(IS-based)}}\\
\cline{2-3}\cline{5-6}\cline{8-9}
&\\[-7pt]
&\!\!\! Micro- \!\!\!&\!\!\! Macro- \!\!\!&\!\!\!\!\!\!&\!\!\! Micro- \!\!\!&\!\!\! Macro- \!\!\!&\!\!\!\!\!\!&\!\!\! Micro- \!\!\!&\!\!\! Macro- \!\!\!\\[-1pt]
\multicolumn{1}{c}{} &\!\!\! Fscore \!\!\!&\!\!\! Fscore \!\!\!&\!\!\!\!\!\!&\!\!\! Fscore \!\!\!&\!\!\! Fscore \!\!\!&\!\!\!\!\!\!&\!\!\! Fscore \!\!\!&\!\!\! Fscore \!\!\!\\
\wcline{1-9}
&\\[-5pt]
\multirow{2}{*}{Strong MTL} \hspace{-8pt}&\!\!\! 91.42\% \!\!\!&\!\!\! 91.68\% \!\!\!&\!\!\!\!\!\!&\!\!\! \textbf{53.91\%} \!\!\!&\!\!\! \textbf{24.09\%} \!\!\!&\!\!\!\!\!\!&\!\!\! \textbf{26.22\%} \!\!\!&\!\!\! \textbf{16.81\%}\!\!\!\\[0pt]
&\!\!\! $\pm$3.00 \!\!\!&\!\!\! \!\!\!\!\!\! $\pm$3.09 \!\!\!\!\!\!&\!\!\!\!\!\!&\!\!\! $\pm${\bf 0.94} \!\!\!&\!\!\! $\pm${\bf 0.83} &\!\!\!\!\!\!& \!\!\!$\pm${\bf 1.50} \!\!\!&\!\!\! $\pm${\bf 1.32}\\
\multirow{2}{*}{Weak MTL} \hspace{-12pt}&\!\!\! 90.51\% \!\!\!&\!\!\! 90.57\% &\!\!\!\!\!\!&\!\!\! 22.74\% \!\!\!&\!\!\! 10.60\% \!\!\!&\!\!\!\!\!\!&\!\!\! 8.66\% \!\!\!&\!\!\! 7.18\% \!\!\!\\[1pt]
\multicolumn{1}{c}{} &\!\!\! $\pm$2.97 \!\!\!&\!\!\! $\pm$3.31 \hspace{-12pt}&\!\!\!\!\!\!&\!\!\! $\pm$18.99 \!\!\!&\!\!\! $\pm$7.37 \!\!\!&\!\!\!\!\!\!&\!\!\! $\pm$1.26 \!\!\!&\!\!\! $\pm$1.00 \!\!\!\\[1pt]
Strong MTL \hspace{-18pt}&\!\!\! 91.78\% \!\!\!&\!\!\! 91.86\% \!\!\!&\!\!\!\!\!\!&\!\!\! 49.01\% \!\!\!&\!\!\! 15.95\% \!\!\!&\!\!\!\!\!\!&\!\!\! 20.90\% \!\!\!&\!\!\! 10.01\% \!\!\!\\[0pt]
w/ reduced data&\!\!\! $\pm$2.19 \!\!\!&\!\!\! $\pm$2.38 \hspace{-12pt}&\!\!\!\!\!\!& \!\!\!$\pm$2.03 \!\!\!&\!\!\! $\pm$1.77 \!\!&\!\!\!\!\!\!&\!\!\! $\pm$2.74 \!\!\!&\!\!\! $\pm$1.88 \!\!\!\\[1pt]
\multirow{1.1}{*}{Semi-MTL} \hspace{-18pt}&\!\!\! 91.76\% \!\!\!&\!\!\! 92.08\% \!\!\!&\!\!\!\!\!\!&\!\!\! 52.11\% \!\!\!&\!\!\! 21.58\% \!\!\!&\!\!\!\!\!\!&\!\!\! 23.57\% \!\!\!&\!\!\! 14.55\% \!\!\!\\[0pt]
\multirow{-1.1}{*}{w/ weak labels} &\!\!\! $\pm$2.70 \!\!\!&\!\!\! $\pm$2.81 \hspace{-12pt}&\!\!\!\!\!\!& \!\!\!$\pm$1.98 \!\!\!&\!\!\! $\pm$1.35 \!\!&\!\!\!\!\!\!&\!\!\! $\pm$2.21 \!\!\!&\!\!\! $\pm$1.75 \!\!\!\\[1pt]
\multirow{1.1}{*}{Semi-MTL} \hspace{-18pt}&\!\!\! \multirow{2}{*}{\textbf{92.12\%}} \!\!\!&\!\!\! \multirow{2}{*}{\textbf{92.58\%}} \!\!\!&\!\!\!\!\!\!&\!\!\! \multirow{2}{*}{51.77\%} \!\!\!&\!\!\! \multirow{2}{*}{21.51\%} \!\!\!&\!\!\!\!\!\!&\!\!\! \multirow{2}{*}{23.96\%} \!\!\!&\!\!\! \multirow{2}{*}{14.87\%} \!\!\!\\[0pt]
w/ partial labels &\!\!\! \multirow{2}{*}{$\pm${\bf 2.59}} \!\!\!&\!\!\! \multirow{2}{*}{$\pm${\bf 2.43}} \hspace{-12pt}&\!\!\!\!\!\!& \!\!\!\multirow{2}{*}{$\pm$1.76} \!\!\!&\!\!\! \multirow{2}{*}{$\pm$1.24} \!\!&\!\!\!\!\!\!&\!\!\! \multirow{2}{*}{$\pm$1.69} \!\!\!&\!\!\! \multirow{2}{*}{$\pm$1.47} \!\!\!\\[0pt]
(proposed) &\!\!\! \!\!\!&\!\!\! \!\!\!&\!\!\! \!\!\!&\!\!\! \!\!\!&\!\!\! \!\!&\!\!\! \!\!\!&\!\!\! \!\!\!&\!\!\! \!\!\!\\
\wcline{1-9}
\end{tabular}
\vspace{10pt}
\end{table}
%
%
%- - - - - - - - - - - - - - - - - - - - - - - - - - - -
%\vspace{-0pt}
\subsection{Experimental Results}
\label{ssec:result}
%\vspace{-0pt}
%- - - - - - - - - - - - - - - - - - - - - - - - - - - -
%-  -  -  -  -  -  -  -  -  -  -  -  -  -  -  -  -  -  -  -
%\vspace{0pt}
\subsubsection{Overall performance characteristics of ASC and SED}
\label{sssec:result1}
%\vspace{0pt}
%-  -  -  -  -  -  -  -  -  -  -  -  -  -  -  -  -  -  -  -
Table~\ref{tbl:performance01} shows the overall performance of ASC and SED in terms of Fscore, especially in the segment-based and intersection-based (IS-based) metrics (\cite{Bilen_ICASSP2020_01}) for SED.
In our experiments, we refer to the methods using strong and weak labels of sound events as strong MTL and weak MTL, respectively.
The semi-supervised methods using weak and partial labels are referred to as semi-MTL w/ weak labels and semi-MTL w/ partial labels, respectively.
For the semi-MTL conditions, we conducted the experiments using 30\% of the strongly labeled data and 70\% of the weakly/partially labeled data.
We also conducted experiments under a condition where data without strong labels were excluded from training.
This setting is referred to as Strong MTL w/ reduced data.

The results show that the semi-MTL-based methods achieve reasonable micro- and macro-Fscores for ASC that are similar to those of the conventional strong and weak MTL methods.
In particular, the proposed semi-supervised approach using partial labels outperformed conventional MTL methods in ASC.
This is because the partial labels for sound events, which were generated using acoustic scene labels from an LLM, contain information on acoustic scenes, and they may have enhanced scene classification.

For SED, the proposed semi-supervised methods with partial labels achieves the detection performance equivalent to that of the conventional semi-supervised method using weak labels in terms of both segment- and IS-based metrics.
This result indicates that the proposed method can further reduce the annotation costs for sound events compared to the conventional semi-supervised method with promising SED results.
\begin{figure}[t!]
\centering
%\hspace*{-3pt}
%\vspace{0pt}
\includegraphics[width=1.0\columnwidth]{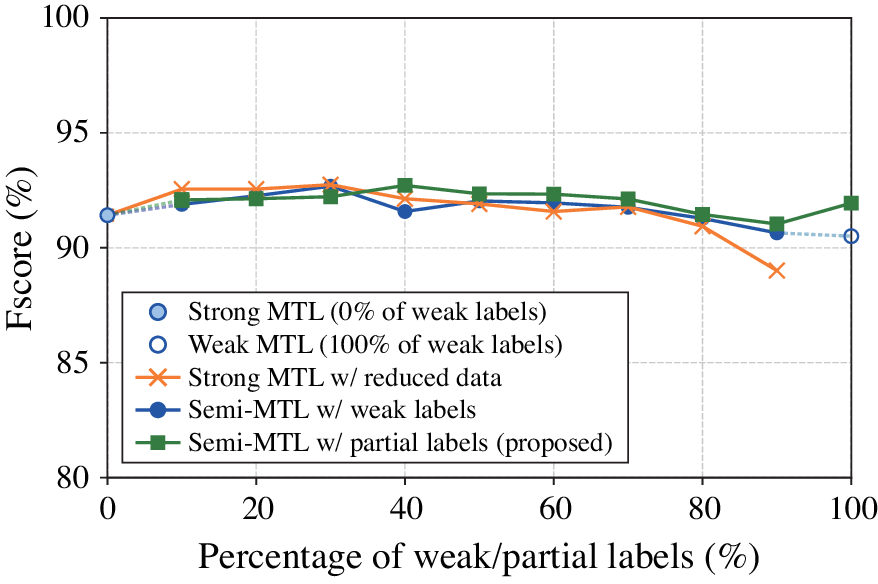}
%\vspace{-12pt}
\caption{ASC performance for various ratios of weakly/partially labeled data of sound events in terms of micro-Fscore}
\label{fig:scene_result_01}
\vspace{10pt}
\end{figure}
%
%
%-  -  -  -  -  -  -  -  -  -  -  -  -  -  -  -  -  -  -  -
%\vspace{0pt}
\subsubsection{Performance characteristics of ASC and SED at various proportion of weak/partial labels}
\label{sssec:result1}
\begin{figure}[t!]
\centering
%\hspace*{-3pt}
%\vspace{0pt}
\includegraphics[width=1.0\columnwidth]{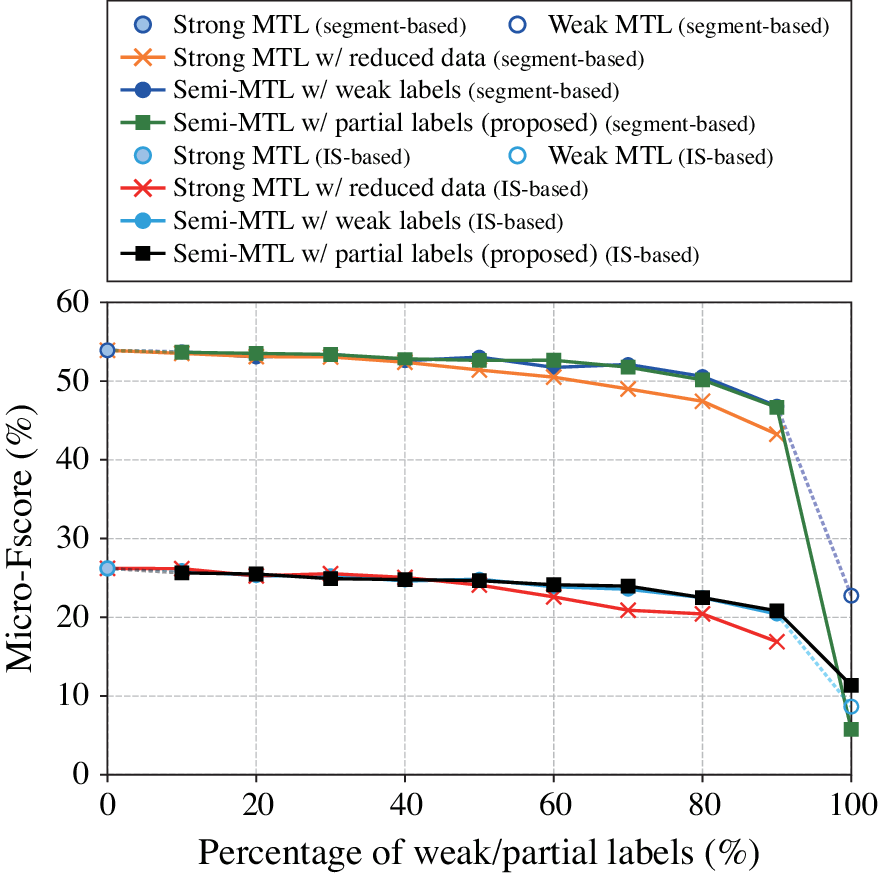}
\vspace{-5pt}
\caption{SED performance for various ratios of weakly/partially labeled data of sound events in terms of micro-Fscore}
\label{fig:event_micro_result_01}
\vspace{10pt}
\end{figure}

\begin{figure}[t!]
\centering
%\vspace{0pt}
\includegraphics[width=1.0\columnwidth]{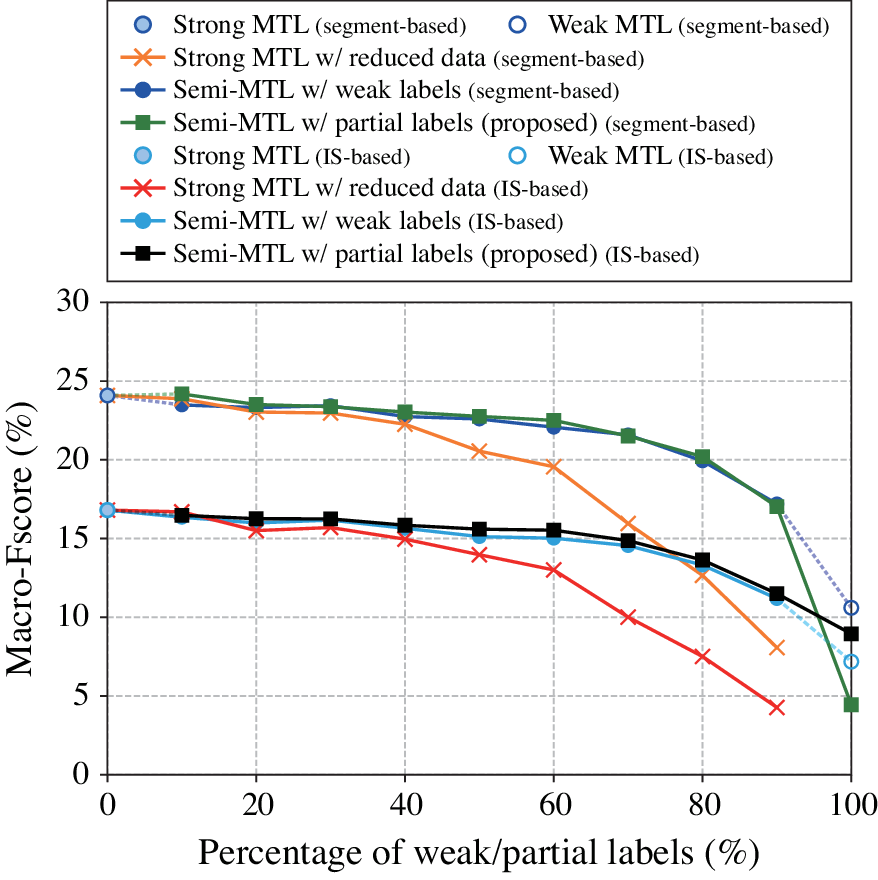}
\vspace{-5pt}
\caption{SED performance for various ratios of weakly/partially labeled data of sound events in terms of macro-Fscore}
\label{fig:event_macro_result_01}
\vspace{3pt}
\end{figure}

%\vspace{0pt}
%-  -  -  -  -  -  -  -  -  -  -  -  -  -  -  -  -  -  -  -
To investigate the detailed behavior of strong, weak, and semi-MTL approaches, we show the evaluation performance of ASC and SED as the proportion of strongly labeled sound event data varies in Figs.~\ref{fig:scene_result_01}--\ref{fig:event_macro_result_01}.
Figure~\ref{fig:scene_result_01} shows that the ASC performance of the proposed semi-MTL approaches remains nearly equivalent to that of the strong MTL approach, even as the proportion of weak/partial labels increases.
This result indicates that ASC does not necessarily require temporal information on sound events, but requires only clip-level information on sound events in acoustic signals.

For the SED performance, Figs.~\ref{fig:event_micro_result_01}--\ref{fig:event_macro_result_01} show that the F-score does not decrease considerably until the proportion of partial labels reaches around 60--70\%.
This result indicates that the proposed semi-MTL approach deliver reasonable performance even when only a small number of strongly labeled data are available alongside a large number of partially labeled data.
Consequently, the proposed methods alleviate the challenges associated with annotating sound event labels.
When comparing the method based on the semi-supervised MTL using weak labels with that using partial labels, we observed nearly equivalent performance in both these methods except when all the training data have weak or partial labels.
This suggests that if part of audio data for the model training do not have strong labels, generating partial labels using LLMs instead of annotating weak labels would be a reasonable solution.
On the other hand, when all training data consist of weak partial labels, the SED performance can degrade significantly.
This result suggests that incorporating strong labels with partial labels and applying semi-supervised learning can substantially enhance the reliability of detection results.

Furthermore, since the results of the proposed method are comparable to those of the Semi-MTL w/ weak labels, it implies that the proposed method remains effective even when using partial labels of the quality shown in Table~\ref{tbl:event_occurrence_partial}.
Thus, the SED performance of the proposed method is comparable across reasonable variations in the size of the sound event label set between that of the actual weak label and the current partial label sets, suggesting that the proposed method is robust to the size of the partial label set.
%
%
%-  -  -  -  -  -  -  -  -  -  -  -  -  -  -  -  -  -  -  -
%\vspace{-0pt}
\subsubsection{Detailed performance evaluation for each acoustic scene and sound event}
\label{sssec:result2}
%\vspace{-0pt}
%-  -  -  -  -  -  -  -  -  -  -  -  -  -  -  -  -  -  -  -
Table~\ref{tab:performance02} shows the detailed ASC performance for each acoustic scene.
The result indicates that there are no significant differences in ASC performance among the strong and semi-MTL approaches.
This also implies that the temporal information on sound events is not critical for each scene classification, and that clip-level sound event information is sufficient for ASC.

Table~\ref{tab:performance03} presents the SED performance and sound duration for each sound event.
These results indicate that the proposed semi-supervised MTL approach using partial labels achieves comparable performance to the method using weak labels in detecting each sound event.
Furthermore, for sound events with longer durations, such as \textit{bird singing}, \textit{fan}, and \textit{large vehicle}, the SED performance is comparable to that of strong MTL.
However, for sound events with short duration, such as \textit{cutlery} and \textit{keyboard typing}, the performance of the proposed method slightly degrades compared with strong MTL.
It is known that the SED model trained with strong labels tends to fail to detect short-duration events compared with that trained with weak labels (\cite{Imoto_AppliedAcoustics2022_01,Igarashi_APSIPA2023_01}).

\begin{table}[t!]
%\vspace{0pt}
\small
\hspace*{-5pt}
\caption{ASC performance for each scene in terms of Fscore}
\label{tab:performance02}
\centering
\begin{tabular}{cccccc}
\wcline{1-6}
& & & & &\\[-7pt]
& {\bf city} & & &{\bf  residential}\\[-2pt]
\multicolumn{1}{c}{\multirow{-1.6}{*}{\bf Method}}&{\bf center}&\multirow{-1.6}{*}{\bf home}&\multirow{-1.6}{*}{\bf office}&{\bf area}\\
\wcline{1-6}\\[-5.5pt]
\multirow{2}{*}{Strong MTL}&91.38\%&94.23\%&96.54\%&84.57\%\\[0pt]
&$\pm$2.38&$\pm$3.47&$\pm$1.85&$\pm$6.28\\[2pt]
\multirow{2}{*}{Weak MTL}&90.45\%&93.40\%&94.16\%&81.95\%\\[0pt]
&$\pm$2.57&$\pm$4.91&$\pm$3.22&$\pm$10.27\\[2pt]
Strong MTL&{\bf 92.27\%}&93.98\%&{\bf 96.63\%}&84.56\%\\[0pt]
w/ reduced data&$\pm${\bf 2.23}&$\pm$2.73&$\pm${\bf 1.69}&$\pm$5.75\\[2pt]
Semi-MTL&90.71\%&95.77\%&95.63\%&86.20\%\\[0pt]
w/ weak labels&$\pm$2.39&$\pm$3.09&$\pm$2.25&$\pm$5.82\\[2pt]
Semi-MTL&\multirow{2}{*}{90.40\%}&\multirow{2}{*}{{\bf 96.44\%}}&\multirow{2}{*}
{96.52\%}&\multirow{2}{*}{{\bf 86.97\%}}\\[0pt]
w/ partial labels&\multirow{2}{*}{$\pm$4.13}&\multirow{2}{*}{$\pm${\bf 1.41}}&\multirow{2}{*}{$\pm$1.88}&\multirow{2}{*}{$\pm${\bf 4.01}}\\[0pt]
(proposed)&&&&\\
\wcline{1-6}
\end{tabular}
\vspace{8pt}
\end{table}

To further investigate this result, Table~\ref{tab:performance04} shows the numbers of true positives (\# TP), false positives (\# FP), and false negatives (\# FN) for each sound event.
Although the proposed method achieves improvements in TP count, it also exhibits an increase in FP count.
This indicates that the proposed method tends to be overconfident in the detection of sound events.
We attribute the overconfidence to confirmation bias (\cite{Arazo_arXiv2019_01}), which reinforces errors in pseudo labels through iterative label refining.
In our proposed method, the partial labels are self-distillated and the MTL model is retrained using the distillated labels. This procedure tends to amplify confidence of the distillated labels.
In particular, prior work (\cite{Arazo_arXiv2019_01}) pointed out that confirmation bias becomes more serious near detection boundaries. 
For SED, short duration events tend to contain many boundary frames relative to their total frames.
As a result, the proposed method result in more TP and FP counts for short duration classes.

%
%\begin{table}[t!]
\begin{sidewaystable}[h]
\small
\centering
\caption{SED performance for each sound event in terms of segment-based Fscore and sound duration. We conducted the experiments with 30\% of the strongly labeled data and 70\% of the weakly/partially labeled data under the semi-MTL condition.}
\vspace{5pt}
\label{tab:performance03}
\begin{tabular}{cccccccccccccc}
\wcline{1-13}
 & & & & & & & & & & & & &\\[-8pt]
&{\bf  bird}&{\bf brakes}& & & & &{\bf glass}&{\bf keyboard}&{\bf large}&{\bf mouse}&{\bf people}&{\bf washing}\\[-1pt]
\multicolumn{1}{c}{\multirow{-1.6}{*}{\!\!\bf Method\!\!}}&\!\!{\bf singing}\!\!&\!\!{\bf squeaking}\!\!&\multirow{-1.6}{*}{\!\!\bf car\!\!}&\multirow{-1.6}{*}{\!\!\bf cutlery\!\!}&\multirow{-1.6}{*}{\!\!\bf dishes\!\!}&\multirow{-1.6}{*}{\!\!\bf fan\!\!}&\!\!{\bf jingling}\!\!&\!\!{\bf typing}\!\!&\!\!{\bf vehicle}\!\!&\!\!{\bf clicking}\!\!&\!\!{\bf walking}\!\!&\!\!{\bf dishes}\!\!\\[1.0pt]
\wcline{1-13}\\[-8pt]
\multirow{2}{*}{Strong MTL} \hspace{-4pt}&40.74\%&{\bf 51.64\%}&{\bf 51.80\%}&{\bf 12.22\%}&12.43\%&{\bf 97.03\%}&0.27\%&{\bf 56.20\%}&17.39\%&{\bf 71.30\%}&{\bf 19.35\%}&5.46\%\\[0pt]
& $\pm$4.48 \!\!\!&\!\!\! $\pm${\bf 5.31} \!\!\!&\!\!\! $\pm${\bf 2.04} \!\!\!&\!\!\! $\pm${\bf 7.90} \!\!\!&\!\!\! $\pm$4.77 \!\!\!&\!\!\! $\pm${\bf 1.20} \!\!\!&\!\!\! $\pm$0.71 \!\!\!&\!\!\! $\pm${\bf 3.52} \!\!\!&\!\!\! $\pm$1.54 \!\!\!&\!\!\! $\pm${\bf 1.86} \!\!\!&\!\!\! $\pm${\bf 3.01} \!\!\!&\!\!\! $\pm$4.34\\[2pt]
\multirow{2}{*}{Weak MTL} \hspace{-4pt}&28.46\%&21.97\%&30.73\%&10.28\%&8.38\%&50.47\%&{\bf 3.57\%}&10.76\%&9.75\%&1.69\%&11.34\%&10.86\%\\[0pt]
& $\pm$17.61 \!\!\!&\!\!\! $\pm$19.38 \!\!\!&\!\!\! $\pm$16.26 \!\!\!&\!\!\! $\pm$9.55 \!\!\!&\!\!\! $\pm$6.27 \!\!\!&\!\!\! $\pm$45.46 \!\!\!&\!\!\! $\pm${\bf 3.84} \!\!\!&\!\!\! $\pm$10.29 \!\!\!&\!\!\! $\pm$6.21 \!\!\!&\!\!\! $\pm$1.83 \!\!\!&\!\!\! $\pm$2.77 \!\!\!&\!\!\! $\pm$10.48\\[2pt]
Strong MTL \hspace{-4pt}&37.08\%&8.59\%&49.35\%&0.08\%&3.52\%&95.17\%&0.00\%&46.23\%&18.80\%&27.68\%&14.01\%&12.49\%\\[0pt]
w/ reduced data& $\pm$8.15 \!\!\!&\!\!\! $\pm$11.03 \!\!\!&\!\!\! $\pm$4.19 \!\!\!&\!\!\! $\pm$0.36 \!\!\!&\!\!\! $\pm$4.84 \!\!\!&\!\!\! $\pm$2.03 \!\!\!&\!\!\! $\pm$0.00 \!\!\!&\!\!\! $\pm$9.11 \!\!\!&\!\!\! $\pm$3.11 \!\!\!&\!\!\! $\pm$22.40 \!\!\!&\!\!\! $\pm$4.97 \!\!\!&\!\!\! $\pm$12.95\\[2pt]
Semi-MTL \hspace{-4pt}&39.94\%&34.44\%&49.97\%&8.12\%&{\bf 13.77\%}&95.74\%&0.36\%&52.48\%&{\bf 19.42\%}&67.63\%&17.22\%&{\bf 12.70\%}\\[0pt]
w/ weak labels& $\pm$6.64 \!\!\!&\!\!\! $\pm$15.14 \!\!\!&\!\!\! $\pm$2.85 \!\!\!&\!\!\! $\pm$9.17 \!\!\!&\!\!\! $\pm${\bf 7.45} \!\!\!&\!\!\! $\pm$2.44 \!\!\!&\!\!\! $\pm$1.00 \!\!\!&\!\!\! $\pm$4.21 \!\!\!&\!\!\! $\pm${\bf 4.24} \!\!\!&\!\!\! $\pm$5.09 \!\!\!&\!\!\! $\pm$5.27 \!\!\!&\!\!\! $\pm${\bf 11.43}\\[2pt]
Semi-MTL \hspace{-4pt}&\multirow{2}{*}{{\bf 42.08\%}}&\multirow{2}{*}{37.77\%}&\multirow{2}{*}{50.60\%}&\multirow{2}{*}{4.83\%}&\multirow{2}{*}{11.21\%}&\multirow{2}{*}{96.50\%}&\multirow{2}{*}{0.81\%}&\multirow{2}{*}{50.90\%}&\multirow{2}{*}{18.09\%}&\multirow{2}{*}{69.15\%}&\multirow{2}{*}{17.41\%}&\multirow{2}{*}{11.23\%}\\[0pt]
w/ partial labels& \multirow{2}{*}{$\pm${\bf 7.85}} \!\!\!&\!\!\! \multirow{2}{*}{$\pm$11.91} \!\!\!&\!\!\! \multirow{2}{*}{$\pm$2.48} \!\!\!&\!\!\! \multirow{2}{*}{$\pm$5.93} \!\!\!&\!\!\! \multirow{2}{*}{$\pm$5.64} \!\!\!&\!\!\! \multirow{2}{*}{$\pm$1.42} \!\!\!&\!\!\! \multirow{2}{*}{$\pm$1.93} \!\!\!&\!\!\! \multirow{2}{*}{$\pm$6.67} \!\!\!&\!\!\! \multirow{2}{*}{$\pm$3.25} \!\!\!&\!\!\! \multirow{2}{*}{$\pm$2.17} \!\!\!&\!\!\! \multirow{2}{*}{$\pm$5.80} \!\!\!&\!\!\! \multirow{2}{*}{$\pm$12.86}\\[0pt]
(proposed) \hspace{-4pt}&&&&&&&&&&&&\\
\wcline{1-13}
& & & & & & & & & & & & &\\[-8pt]
Average sound\hspace{-4pt}&\!\!\!7.63\!\!\!&\!\!\!1.65\!\!\!&\!\!\!6.88\!\!\!&\!\!\!0.74\!\!\!&\!\!\!1.24\!\!\!&\!\!\!29.99\!\!\!&\!\!\!0.80\!\!\!&\!\!\!0.21\!\!\!&\!\!\!14.68\!\!\!&\!\!\!0.14\!\!\!&\!\!\!6.63\!\!\!&\!\!\!4.15\!\!\!&\\
duration (s)&\!\!\!$\pm$8.49\!\!\!&\!\!\!$\pm$1.97\!\!\!&\!\!\!$\pm$4.72\!\!\!&\!\!\!$\pm$0.53\!\!\!&\!\!\!$\pm$1.12\!\!\!&\!\!\!$\pm$0.01\!\!\!&\!\!\!$\pm$0.46\!\!\!&\!\!\!$\pm$0.22\!\!\!&\!\!\!$\pm$7.35\!\!\!&\!\!\!$\pm$0.08\!\!\!&\!\!\!$\pm$8.78\!\!\!&\!\!\!$\pm$3.75\!\!\!&\\
\wcline{1-13}
\end{tabular}
%\end{table}
\end{sidewaystable}
%
%
%
%\begin{table}[t!]
\begin{sidewaystable}[h]
\small
\centering
\caption{Average numbers of true positive, false positive, and false negative samples for each sound event. We conducted the experiments with 30\% of the strongly labeled data and 70\% of the weakly/partially labeled data under the semi-MTL condition.}
\vspace{5pt}
\label{tab:performance04}
\begin{tabular}{crrrrrrrrrrrrr}
\wcline{1-14}
 & & & & & & & & & & & & &\\[-8pt]
&&\multicolumn{1}{c}{\!\!{\bf bird}\!\!}&\multicolumn{1}{c}{\!\!{\bf brakes}\!\!}& & & & &\multicolumn{1}{c}{\!\!{\bf glass}\!\!}&\multicolumn{1}{c}{\!\!{\bf keyboard}\!\!}&\multicolumn{1}{c}{\!\!{\bf large}\!\!}&\multicolumn{1}{c}{\!\!{\bf mouse}\!\!}&\multicolumn{1}{c}{\!\!{\bf people}\!\!}&\multicolumn{1}{c}{\!\!{\bf washing}\!\!}\\[0pt]
\multicolumn{1}{c}{\multirow{-1.6}{*}{\bf Method}\!\!}&\multicolumn{1}{c}{\multirow{-1.6}{*}{\!\!\bf Metric\!\!}\!\!}&\multicolumn{1}{c}{\!\!{\bf singing}\!\!}&\multicolumn{1}{c}{\!\!{\bf squeaking}\!\!}&\multicolumn{1}{c}{\multirow{-1.6}{*}{\bf car}\!\!}&\multicolumn{1}{c}{\multirow{-1.6}{*}{\bf cutlery}\!\!}&\multicolumn{1}{c}{\multirow{-1.6}{*}{\bf dishes}\!\!}&\multicolumn{1}{c}{\multirow{-1.6}{*}{\bf fan}\!\!}&\multicolumn{1}{c}{\!\!{\bf jingling}\!\!}&\multicolumn{1}{c}{\!\!{\bf typing}\!\!}&\multicolumn{1}{c}{\!\!{\bf vehicle}\!\!}&\multicolumn{1}{c}{\!\!{\bf clicking}\!\!}&\multicolumn{1}{c}{\!\!{\bf walking}\!\!}&\multicolumn{1}{c}{\!\!{\bf dishes}\!\!}\\[0pt]
&&&&&&&&&&&&\\[-8pt]
\wcline{1-14}\\[-8pt]
&\!\!\!\!\!\!\# TP&\!\!\!\!\!\!6,745.1&\!\!\!\!\!\!1,767.0&\!\!20,341.2&\!\!\!\!\!\!67.4&\!\!\!\!\!\!262.4&\!\!\!\!\!37,386.1&\!\!\!\!\!\!0.4&\!\!\!\!\!\!1,131.7&\!\!\!\!\!\!2,822.1&\!\!\!\!\!\!515.6&\!\!\!\!\!\!1,833.7&\!\!\!\!\!\!230.1\\[0pt]
Strong MTL&\!\!\!\!\!\!\# FP&\!\!\!\!\!\!5,643.2&\!\!\!\!\!\!998.4&\!\!24,300.0&\!\!\!\!\!\!53.3&\!\!\!\!\!\!380.5&\!\!\!\!\!745.3&\!\!\!\!\!\!3.8&\!\!\!\!\!\!653.4&\!\!\!\!\!\!23,550.3&\!\!\!\!\!\!125.6&\!\!\!\!\!\!4,550.5&\!\!\!\!\!\!1,039.3\\[0pt]
&\!\!\!\!\!\!\# FN&\!\!\!\!\!\!13,743.0&\!\!\!\!\!\!2,270.1&\!\!13,539.8&\!\!\!\!\!\!877.6&\!\!\!\!\!\!3,203.6&\!\!\!\!\!1,559.9&\!\!\!\!\!\!269.6&\!\!\!\!\!\!1,092.3&\!\!\!\!\!\!3,340.9&\!\!\!\!\!\!289.4&\!\!\!\!\!\!10,683.3&\!\!\!\!\!\!6,310.9\\[2pt]
&\!\!\!\!\!\!\# TP&\!\!\!\!\!\!8,286.1&\!\!\!\!\!\!2,215.8&\!\!27,726.2&\!\!\!\!\!\!70.3&\!\!\!\!\!\!702.8&\!\!\!\!\!37,360.5&\!\!\!\!\!\!2.8&\!\!\!\!\!\!2,149.9&\!\!\!\!\!\!3,267.8&\!\!\!\!\!\!673.6&\!\!\!\!\!\!2,905.1&\!\!\!\!\!\!1,282.8\\[0pt]
Weak MTL&\!\!\!\!\!\!\# FP&\!\!\!\!\!\!10,596.8&\!\!\!\!\!\!12,257.7&\!\!53,787.9&\!\!\!\!\!\!2,205.9&\!\!\!\!\!\!5,680.4&\!\!\!\!\!2,207.6&\!\!\!\!\!\!485.4&\!\!\!\!\!\!23,670.9&\!\!\!\!\!\!32,417.8&\!\!\!\!\!\!24,349.7&\!\!\!\!\!\!24,488.0&\!\!\!\!\!\!4,777.8\\[0pt]
&\!\!\!\!\!\!\# FN&\!\!\!\!\!\!12,202.0&\!\!\!\!\!\!1,821.2&\!\!6,154.8&\!\!\!\!\!\!874.7&\!\!\!\!\!\!2,763.2&\!\!\!\!\!1,585.5&\!\!\!\!\!\!267.2&\!\!\!\!\!\!74.1&\!\!\!\!\!\!2,895.2&\!\!\!\!\!\!131.4&\!\!\!\!\!\!9,612.0&\!\!\!\!\!\!5,258.2
\\[2pt]
\multirow{2}{*}{Strong MTL}&\!\!\!\!\!\!\# TP&\!\!\!\!\!\!6,075.9&\!\!\!\!\!\!204.0&\!\!19,592.0&\!\!\!\!\!\!0.4&\!\!\!\!\!\!80.9&\!\!\!\!\!37,376.7&\!\!\!\!\!\!0.0&\!\!\!\!\!\!983.7&\!\!\!\!\!\!2,858.7&\!\!\!\!\!\!150.4&\!\!\!\!\!\!1,593.5&\!\!\!\!\!\!739.0\\[0pt]
\multirow{2}{*}{w/ reduced data}&\!\!\!\!\!\!\# FP&\!\!\!\!\!\!5,296.1&\!\!\!\!\!\!36.6&\!\!25,261.8&\!\!\!\!\!\!0.2&\!\!\!\!\!\!221.4&\!\!\!\!\!2,256.4&\!\!\!\!\!\!0.0&\!\!\!\!\!\!951.6&\!\!\!\!\!\!22,369.6&\!\!\!\!\!\!11.8&\!\!\!\!\!\!6,594.2&\!\!\!\!\!\!2,191.1\\[0pt]
&\!\!\!\!\!\!\# FN&\!\!\!\!\!\!14,412.1&\!\!\!\!\!\!3,833.0&\!\!14,289.0&\!\!\!\!\!\!944.6&\!\!\!\!\!\!3,385.1&\!\!\!\!\!1,569.3&\!\!\!\!\!\!270.0&\!\!\!\!\!\!1,240.3&\!\!\!\!\!\!3,304.3&\!\!\!\!\!\!654.6&\!\!\!\!\!\!10,923.5&\!\!\!\!\!\!5,802.1
\\[2pt]
\multirow{2}{*}{Semi-MTL}&\!\!\!\!\!\!\# TP&\!\!\!\!\!\!6,679.1&\!\!\!\!\!\!1,001.0&\!\!19,050.3&\!\!\!\!\!\!45.8&\!\!\!\!\!\!332.0&\!\!\!\!\!37,339.6&\!\!\!\!\!\!0.5&\!\!\!\!\!\!1,141.3&\!\!\!\!\!\!2,430.1&\!\!\!\!\!\!470.9&\!\!\!\!\!\!1,641.5&\!\!\!\!\!\!678.4
\\[0pt]
\multirow{2}{*}{w/ weak labels}&\!\!\!\!\!\!\# FP&\!\!\!\!\!\!5,689.3&\!\!\!\!\!\!447.2&\!\!23,169.8&\!\!\!\!\!\!37.7&\!\!\!\!\!\!636.2&\!\!\!\!\!1,769.6&\!\!\!\!\!\!4.1&\!\!\!\!\!\!998.8&\!\!\!\!\!\!17,169.1&\!\!\!\!\!\!111.9&\!\!\!\!\!\!4,688.4&\!\!\!\!\!\!2,101.3
\\[0pt]
&\!\!\!\!\!\!\# FN&\!\!\!\!\!\!13,809.0&\!\!\!\!\!\!3,036.1&\!\!14,830.7&\!\!\!\!\!\!899.2&\!\!\!\!\!\!3,134.1&\!\!\!\!\!1,606.4&\!\!\!\!\!\!269.5&\!\!\!\!\!\!1,082.7&\!\!\!\!\!\!3,732.9&\!\!\!\!\!\!334.1&\!\!\!\!\!\!10,875.5&\!\!\!\!\!\!5,862.6
\\[2pt]
Semi-MTL&\!\!\!\!\!\!\# TP&\!\!\!\!\!\!8,012.2&\!\!\!\!\!\!1,407.2&\!\!24,614.3&\!\!\!\!\!\!57.7&\!\!\!\!\!\!434.6&\!\!\!\!\!37,397.0&\!\!\!\!\!\!1.3&\!\!\!\!\!\!1,369.2&\!\!\!\!\!\!3,081.1&\!\!\!\!\!\!552.6&\!\!\!\!\!\!2,051.9&\!\!\!\!\!\!974.4
\\[0pt]
w/ partial labels&\!\!\!\!\!\!\# FP&\!\!\!\!\!\!7,870.2&\!\!\!\!\!\!1,456.1&\!\!36,710.0&\!\!\!\!\!\!105.0&\!\!\!\!\!\!905.7&\!\!\!\!\!536.5&\!\!\!\!\!\!10.8&\!\!\!\!\!\!1,469.3&\!\!\!\!\!\!27,294.4&\!\!\!\!\!\!268.1&\!\!\!\!\!\!9,644.0&\!\!\!\!\!\!2,711.4
\\[0pt]
(proposed)&\!\!\!\!\!\!\# FN&\!\!\!\!\!\!12,475.8&\!\!\!\!\!\!2,629.8&\!\!9,266.7&\!\!\!\!\!\!887.3&\!\!\!\!\!\!3,031.4&\!\!\!\!\!1,549.1&\!\!\!\!\!\!268.7&\!\!\!\!\!\!854.8&\!\!\!\!\!\!3,082.0&\!\!\!\!\!\!252.4&\!\!\!\!\!\!10,465.1&\!\!\!\!\!\!5,566.6
\\[2pt]
\wcline{1-14}
\end{tabular}
\end{sidewaystable}
\clearpage
\begin{figure}[t!]
\centering
%\hspace*{-3pt}
%\vspace{0pt}
\includegraphics[width=1.0\columnwidth]{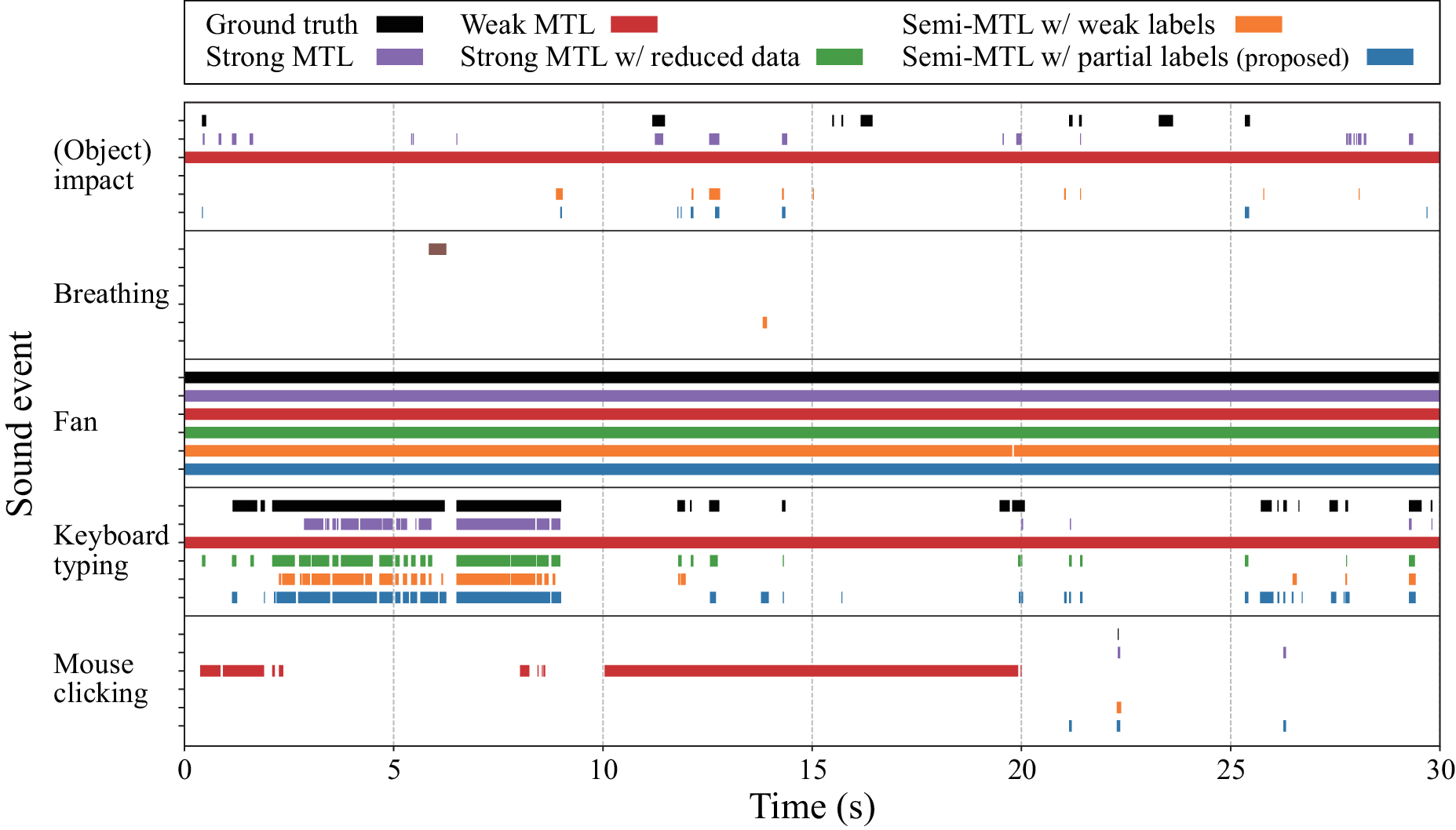}
\vspace{-10pt}
\caption{Sound event detection results for 276.wav recorded in an office scene from the TUT Acoustic Scenes 2016 dataset. Only sound events that include multiple ground truth labels or detected events are shown. We conducted the experiments with 30\% of the strongly labeled data and 70\% of the weakly/partially labeled data under the semi-MTL condition.}
\label{fig:event_plot_01}
%\vspace{0pt}
\end{figure}
%
%
%
%-  -  -  -  -  -  -  -  -  -  -  -  -  -  -  -  -  -  -  -
%\vspace{-0pt}
\subsubsection{Qualitative analysis of sound event detection results}
\label{sssec:Qualitative_analysis}
\begin{figure}[t!]
\centering
%\hspace*{-3pt}
%\vspace{0pt}
\includegraphics[width=1.0\columnwidth]{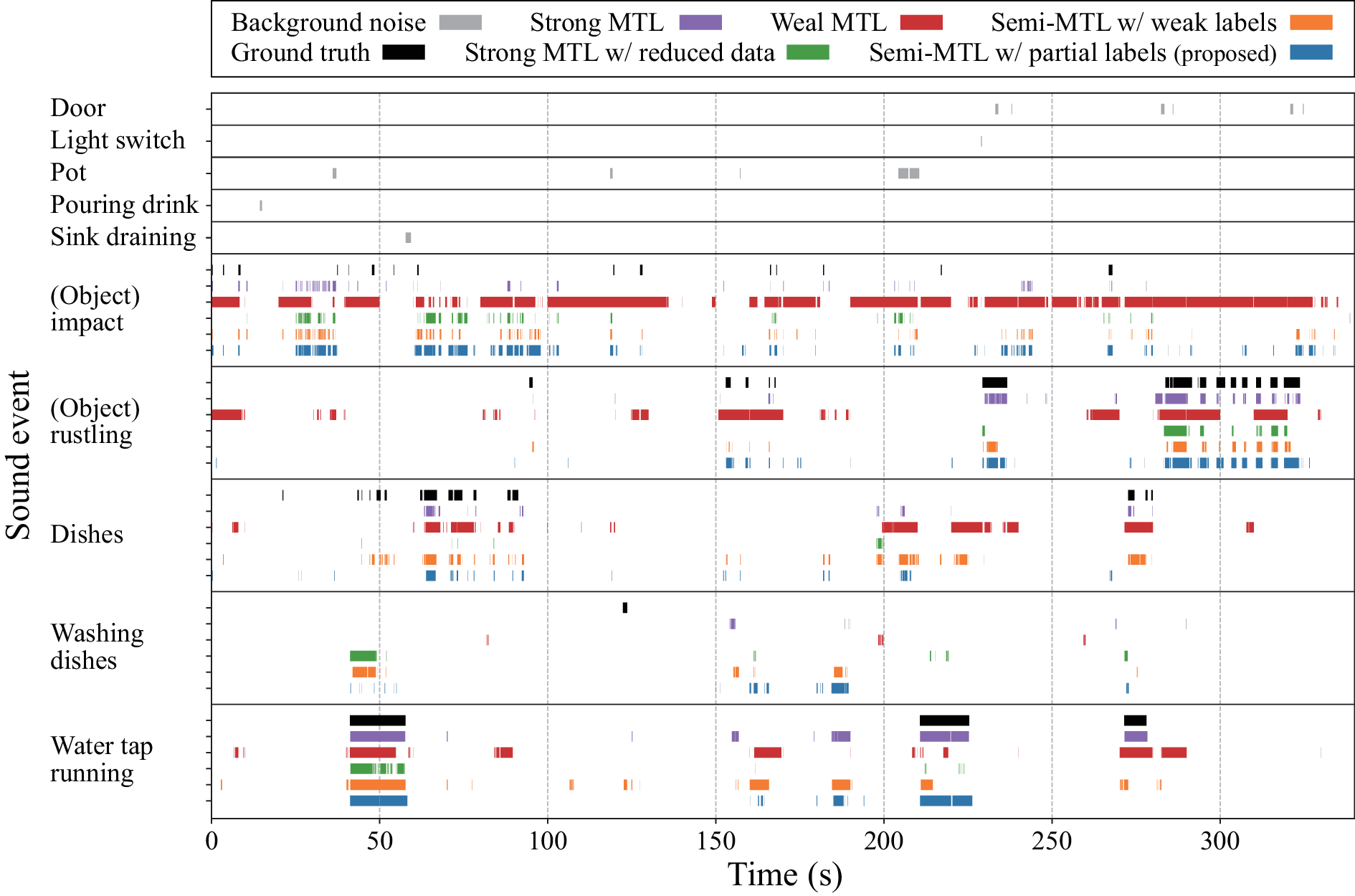}
\vspace{-10pt}
\caption{Sound event detection results for b043.wav recorded in a home scene from the TUT Sound Events 2016 dataset. The figure includes the additional visualization of background noise not annotated in the ground truth. Only sound events that include multiple ground truth labels or detected events are shown. We conducted the experiments with 30\% of the strongly labeled data and 70\% of the weakly/partially labeled data under the semi-MTL condition.}
\label{fig:event_plot_02}
\vspace{-10pt}
%\vspace{0pt}
\end{figure}

%\vspace{-0pt}
%-  -  -  -  -  -  -  -  -  -  -  -  -  -  -  -  -  -  -  -
To qualitatively assess the behavior of the proposed method, Figs.~\ref{fig:event_plot_01} and \ref{fig:event_plot_02} show the detection results on randomly selected sound clips.
Each figure presents the ground truth of sound event labels, the detection outputs from the conventional and proposed methods.

In Fig.~\ref{fig:event_plot_01}, the proposed method shows a more accurate detection performance for the \textit{keyboard typing} event than the conventional strong MTL and semi-MTL methods with weak labels.
In addition, we observed false positives where events were detected at the correct time boundary but with incorrect labels; for example, \textit{keyboard typing} and \textit{mouse clicking} were detected instead of \textit{(object) impact}.
For these cross-triggering cases, incorporating a more refined mechanism for sound event classification may help mitigate such errors.

Figure~\ref{fig:event_plot_02} includes the additional visualization of background noise, which corresponds to sound events not annotated as ground truth labels.
These visualizations enable us to assess the model robustness to background sounds.
The results indicate that the proposed method is as robust as the conventional strong MTL and semi-MTL methods in ignoring irrelevant background noise, and it still can detect target sound events.
%
%
%-  -  -  -  -  -  -  -  -  -  -  -  -  -  -  -  -  -  -  -
%\vspace{-0pt}
\subsubsection{Model complexity and training cost}
\label{sssec:computational_cost}
%\vspace{-0pt}
%-  -  -  -  -  -  -  -  -  -  -  -  -  -  -  -  -  -  -  -
Table~\ref{tab:calculation_cost} shows the numbers of model parameters and training costs for the proposed and conventional methods.
Note that, in the proposed method, the parameters used in the distillation module can be reused within the main module, which eliminates the use of additional model parameters.
As shown in the table, there are no significant differences in the number of model parameters and training time.
This indicates that our proposed method can be implemented without considerably increasing additional computational cost or memory requirements.
\begin{table}[t!]
\small
\caption{Comparison of model size and training cost among proposed and conventional methods. For the semi-MTL conditions, we conducted the experiments using 30\% of the strongly labeled data and 70\% of the weakly/partially labeled data.}
%\vspace{0pt}
\vspace{10pt}
\hspace*{-8pt}
\label{tab:calculation_cost}
\centering
\begin{tabular}{cccc}
\wcline{1-4}
& & &\\[-5.5pt]
&{\bf \# parameters} \!\!\!\!&\!\!\!\! {\bf Training} \!\!\!\!&\!\!\!\! {\bf Training time ratio}\\[-1pt]
\multicolumn{1}{c}{\multirow{-2.0}{*}{\bf Method}}&{\bf (k)}&{\bf time (s)}\!\!\!\!&\!\!\!\!{\bf (Strong MTL = 1.0)}\\
\wcline{1-4}\\[-5.5pt]
Strong MTL&1,323&314.8 $\pm$ 2.2&1.00\\[2pt]
Weak MTL&1,323&311.9 $\pm$ 1.3&0.99\\[2pt]
Strong MTL&\multirow{2}{*}{1,323}&118.8 $\pm$ 1.0&0.37\\[0pt]
w/ reduced data&&&\\[2pt]
Semi-MTL&\multirow{2}{*}{1,331}&327.1 $\pm$ 1.9&1.04\\[0pt]
w/ weak labels&&&\\[2pt]
Semi-MTL&&&\\[0pt]
w/ partial labels&1,331&340.2 $\pm$ 2.5&1.08\\[0pt]
(proposed)&&&\\
\wcline{1-4}
\end{tabular}
%\vspace{0pt}
\end{table}
%
%
%---------------------------------------------------
%\vspace{-0pt}
\section{Conclusions}
\label{sec:conclusion}
%\vspace{-0pt}
%---------------------------------------------------
We proposed the method for the joint analysis of acoustic scenes and sound events based on the semi-supervised SED strategy using partial labels of sound events.
We further introduced the LLM-based label creation and self-distillation-based label refining methods for the proposed partial label learning in SED.
The results of experiments using our constructed dataset show that the semi-supervised approach using partial labels achieve reasonable performance even with a small number of strongly labeled data and a large number of partially labeled data.
Future work should focus on exploring more effective approaches to refining partial labels of sound events.
Also, the application of partial label learning to single-task SED settings where acoustic scene labels are not available should be addressed.
This will require new strategies for generating candidate event label sets without scene context, which poses a more challenging and general problem.
%
%
%
%\appendices
%---------------------------------------------------
%\vspace{-0pt}
\section*{Appendix: Prompts used to generate partial labels of sound events}
%\label{sec:appendix}
%\vspace{-0pt}
%---------------------------------------------------
To generate partial labels of sound events, we utilized the ChatGPT o3-mini-high on February 02, 2025.
The input prompts used to generate partial labels are shown in Table~\ref{tab:prompt}, which includes the possible sound events, the supplemental explanation of a sound event class, the instruction to consider the partial labels of sound events for each scene, and the output format.
We obtain partial labels and the reasons for including the sound events in the list.
The lists of partial labels and reasons are available\footnote{\url{https://github.com/KeisukeImoto/SED_ASC_partial_label.git}}.

\begin{table}[t!]
\small
\centering
\caption{Prompts for generating partial labels of sound events input into ChatGPT o3-mini-high}
\label{tab:prompt}
\begin{tabularx}{\linewidth}{|X|}
\hline
\ \\[-8pt]
%-------------------------------
Here is the list of 25 possible sound events:\\
object banging, object impact, object rustling, object snapping, object squeaking, bird singing, brakes squeaking, breathing, car, children, cupboard, cutlery, dishes, drawer, fan, glass jingling, keyboard typing, large vehicle, mouse clicking, mouse wheeling, people talking, people walking, washing dishes, water tap running, wind blowing.

Here, ``object'' refers to an unknown sound source, although we can understand how the sound is produced. We can include these ambiguous object sounds in the list.

If we are in a <scene name> scene, which sound events are likely to be heard?
Please list all the sound events one by one (without merging) in CSV format, and provide your reasoning process in a two-column CSV format.\\[2pt]
%-------------------------------
\hline
\end{tabularx}
%\vspace{0pt}
\end{table}
%
%-------------------------------
\section{Acknowledgment}
%-------------------------------
This work was supported by JSPS KAKENHI Grant Numbers 22H03639, 23K16908, and 25H01142.

%\clearpage
\bibliographystyle{IEEEtran}
%\begin{small}
\bibliography{IEEEabrv.bib,KeisukeImoto13.bib,ATSIP2025refs.bib}

\begin{thebibliography}{10}
\providecommand{\url}[1]{#1}
\csname url@samestyle\endcsname
\providecommand{\newblock}{\relax}
\providecommand{\bibinfo}[2]{#2}
\providecommand{\BIBentrySTDinterwordspacing}{\spaceskip=0pt\relax}
\providecommand{\BIBentryALTinterwordstretchfactor}{4}
\providecommand{\BIBentryALTinterwordspacing}{\spaceskip=\fontdimen2\font plus
\BIBentryALTinterwordstretchfactor\fontdimen3\font minus
  \fontdimen4\font\relax}
\providecommand{\BIBforeignlanguage}[2]{{%
\expandafter\ifx\csname l@#1\endcsname\relax
\typeout{** WARNING: IEEEtran.bst: No hyphenation pattern has been}%
\typeout{** loaded for the language `#1'. Using the pattern for}%
\typeout{** the default language instead.}%
\else
\language=\csname l@#1\endcsname
\fi
#2}}
\providecommand{\BIBdecl}{\relax}
\BIBdecl

\bibitem{Fonseca_DCASE2018_01}
E.~Fonseca, M.~Plakal, F.~Font, D.~P.~W. Ellis, X.~Favory, J.~Jordi, and
  X.~Serra, ``General-purpose tagging of freesound audio with {AudioSet}
  labels: Task description, dataset, and baseline,'' \emph{Proc. Workshop on
  Detection and Classification of Acoustic Scenes and Events {\rm (}DCASE{\rm
  )}}, pp. 69--73, 2018.

\bibitem{Nishida_EUSIPCO2022_01}
T.~Nishida, K.~Dohi, T.~Endo, M.~Yamamoto, and Y.~Kawaguchi, ``Anomalous sound
  detection based on machine activity detection,'' \emph{Proc. European Signal
  Processing Conference {\rm (}EUSIPCO{\rm )}}, pp. 269--273, 2022.

\bibitem{Morfi_JASA2021_01}
V.~Morfi, R.~F. Lachlan, and D.~Stowell, ``Deep perceptual embeddings for
  unlabelled animal sound,'' \emph{{IEEE/ACM} Trans. Audio Speech Lang.
  Process.}, vol. 150, no.~1, pp. 2--11, 2020.

\bibitem{Valenti_IJCNN2017_01}
M.~Valenti, S.~Squartini, A.~Diment, G.~Parascandolo, and T.~Virtanen, ``A
  convolutional neural network approach for acoustic scene classification,''
  \emph{Proc. International Joint Conference on Neural Networks {\rm
  (}IJCNN{\rm )}}, pp. 1547--1554, 2017.

\bibitem{Ford_INTERSPEECH2019_01}
L.~Ford, H.~Tang, F.~Grondin, and J.~Glass, ``A deep residual network for
  large-scale acoustic scene analysis,'' \emph{Proc. {INTERSPEECH}}, pp.
  2568--2572, 2019.

\bibitem{Kong_TASLP2020_01}
Q.~Kong, Y.~Cao, T.~Iqbal, Y.~Wang, W.~Wang, and M.~D. Plumbley, ``{PANNs}:
  Large-scale pretrained audio neural networks for audio pattern recognition,''
  \emph{{IEEE/ACM} Trans. Audio Speech Lang. Process.}, vol.~28, pp.
  2880--2894, 2020.

\bibitem{Cakir_TASLP2017_01}
E.~\c{C}ak\i r, G.~Parascandolo, T.~Heittola, H.~Huttunen, and T.~Virtanen,
  ``Convolutional recurrent neural networks for polyphonic sound event
  detection,'' \emph{{IEEE/ACM} Trans. Audio Speech Lang. Process.}, vol.~25,
  no.~6, pp. 1291--1303, 2017.

\bibitem{Kong_TASLP2020_02}
Q.~Kong, Y.~Xu, W.~Wang, and M.~D. Plumbley, ``Sound event detection of weakly
  labelled data with {CNN}-{T}ransformer and automatic threshold
  optimization,'' \emph{{IEEE/ACM} Trans. Audio Speech Lang. Process.},
  vol.~28, pp. 2450--2460, 2020.

\bibitem{Miyazaki_DCASE2020_01}
K.~Miyazaki, T.~Komatsu, T.~Hayashi, S.~Watanabe, T.~Toda, and K.~Takeda,
  ``Conformer-based sound event detection with semi-supervised learning and
  data augmentation,'' \emph{Proc. Workshop on Detection and Classification of
  Acoustic Scenes and Events {\rm (}DCASE{\rm )}}, pp. 100--104, 2020.

\bibitem{Mesaros_EUSIPCO2011_01}
A.~Mesaros, T.~Heittola, and A.~Klapuri, ``Latent semantic analysis in sound
  event detection,'' \emph{Proc. European Signal Processing Conference {\rm
  (}EUSIPCO{\rm )}}, pp. 1307--1311, 2011.

\bibitem{Imoto_TASLP2019_01}
K.~Imoto and N.~{Ono}, ``Acoustic topic model for scene analysis with
  intermittently missing observations,'' \emph{{IEEE/ACM} Trans. Audio Speech
  Lang. Process.}, vol.~27, no.~2, pp. 367--382, 2019.

\bibitem{Hou_SLP2023_01}
Y.~Hou, S.~Song, C.~Yu, W.~Wang, and D.~Botteldooren, ``Audio event-relational
  graph representation learning for acoustic scene classification,''
  \emph{{IEEE} Signal Processing Letters}, pp. 1--5, 2023.

\bibitem{Bear_INTERSPEECH2019_01}
H.~L. Bear, I.~Nolasco, and E.~Benetos, ``Towards joint sound scene and
  polyphonic sound event recognition,'' \emph{Proc. {INTERSPEECH}}, pp.
  4594--4598, 2019.

\bibitem{Tonami_WASPAA2019_01}
N.~Tonami, K.~Imoto, M.~Niitsuma, R.~Yamanishi, and Y.~Yamashita, ``Joint
  analysis of acoustic events and scenes based on multitask learning,''
  \emph{Proc. {IEEE} Workshop on Applications of Signal Processing to Audio and
  Acoustics {\rm (}WASPAA{\rm )}}, pp. 333--337, 2019.

\bibitem{Jung_ICASSP2021_01}
J.~Jung, H.~J. Shim, J.~H. Kim, and H.~J. Yu, ``{DCASENET}: An integrated
  pretrained deep neural network for detecting and classifying acoustic scenes
  and events,'' \emph{Proc. {IEEE} International Conference on Acoustics,
  Speech and Signal Processing {\rm (}ICASSP{\rm )}}, pp. 621--625, 2021.

\bibitem{Kumar_ACMMM2016_01}
A.~Kumar and B.~Raj, ``Audio event detection using weakly labeled data,''
  \emph{Proc. {ACM} International Conference on Multimedia {\rm (}ACMMM{\rm
  )}}, pp. 1038--1047, 2016.

\bibitem{Turpault_DCASE2019_01}
N.~Turpault, R.~Serizel, A.~Parag~Shah, and J.~Salamon, ``{Sound Event
  Detection in Domestic Environments with Weakly Labeled Data and Soundscape
  Synthesis},'' \emph{Proc. Workshop on Detection and Classification of
  Acoustic Scenes and Events {\rm (}DCASE{\rm )}}, pp. 253--257, 2019.

\bibitem{Tsubaki_IWAENC2022_01}
S.~Tsubaki, K.~Imoto, and N.~Ono, ``Joint analysis of acoustic scenes and sound
  events with weakly labeled data,'' \emph{Proc. International Workshop on
  Acoustic Signal Enhancement {\rm (}IWAENC{\rm )}}, pp. 1--5, 2022.

\bibitem{Igarashi_APSIPA2023_01}
A.~Igarashi, S.~Tsubaki, D.~Niizumi, D.~Takeuchi, N.~Harada, and K.~Imoto,
  ``Joint analysis of acoustic scenes and sound events based on semi-supervised
  approach,'' \emph{Proc. Asia-Pacific Signal and Information Processing
  Association Annual Summit and Conference {\rm (}APSIPA ASC{\rm )}}, pp.
  2050--2056, 2023.

\bibitem{Cour_JMLR2011_01}
T.~Cour, B.~Sapp, and B.~Taskar, ``Learning from partial labels,''
  \emph{Journal of Machine Learning Research}, vol.~12, no.~42, pp. 1501--1536,
  2011.

\bibitem{Dietterich_AI1997_01}
T.~G. Dietterich, R.~H. Lathrop, and T.~Lozano-P\'{e}rez, ``Solving the
  multiple instance problem with axis-parallel rectangles,'' \emph{Artificial
  Intelligence}, vol.~89, no.~1, pp. 31--71, 1997.

\bibitem{Wang_ICASSP2019_01}
Y.~Wang, J.~Li, and F.~Metze, ``A comparison of five multiple instance learning
  pooling functions for sound event detection with weak labeling,'' \emph{Proc.
  {IEEE} International Conference on Acoustics, Speech and Signal Processing
  {\rm (}ICASSP{\rm )}}, pp. 31--35, 2019.

\bibitem{Liu_ICLR2020_01}
L.~Liu, H.~Jiang, P.~He, W.~Chen, X.~Liu, J.~Gao, and J.~Han, ``On the variance
  of the adaptive learning rate and beyond,'' \emph{Proc. International
  Conference on Learning Representations {\rm (}ICLR{\rm )}}, pp. 1--13, 2020.

\bibitem{Mesaros_EUSIPCO2016_01}
A.~Mesaros, T.~Heittola, and T.~Virtanen, ``{TUT} database for acoustic scene
  classification and sound event detection,'' \emph{Proc. European Signal
  Processing Conference {\rm (}EUSIPCO{\rm )}}, pp. 1128--1132, 2016.

\bibitem{Mesaros_DCASE2017_01}
A.~Mesaros, T.~Heittola, A.~Diment, B.~Elizalde, A.~Shah, B.~Raj, and
  T.~Virtanen, ``{DCASE} 2017 challenge setup: Tasks, datasets and baseline
  system,'' \emph{Proc. Workshop on Detection and Classification of Acoustic
  Scenes and Events {\rm (}DCASE{\rm )}}, pp. 85--92, 2017.

\bibitem{Bilen_ICASSP2020_01}
C.~Bilen, G.~Ferroni, F.~Tuveri, J.~Azcarreta, and S.~Krstulovic, ``A framework
  for the robust evaluation of sound event detection,'' \emph{Proc. {IEEE}
  International Conference on Acoustics, Speech and Signal Processing {\rm
  (}ICASSP{\rm )}}, pp. 61--65, 2020.

\bibitem{Imoto_AppliedAcoustics2022_01}
K.~Imoto, S.~Mishima, Y.~Arai, and R.~Kondo, ``Impact of data imbalance caused
  by inactive frames and difference in sound duration on sound event detection
  performance,'' \emph{Applied Acoustics}, vol. 196, no. 108882, 2022.

\bibitem{Arazo_arXiv2019_01}
E.~Arazo, D.~Ortego, P.~Albert, N.~E. O'Connor, and K.~McGuinness,
  ``Pseudo-labeling and confirmation bias in deep semi-supervised learning,''
  \emph{arXiv, arXiv:1908.02983}, 2019.

\end{thebibliography}

%\end{small}
\end{document}